\documentclass[a4paper]{report}
\usepackage[utf8]{inputenc}
\usepackage[T1]{fontenc}
\usepackage{RJournal}
\usepackage{amsmath,amssymb,array}
\usepackage{booktabs}

\usepackage{longtable}

\makeatletter
 \let\@cite@ofmt\@firstofone
 \def\@biblabel#1{}
 \def\@cite#1#2{{#1\if@tempswa , #2\fi}}
\makeatother
\newlength{\cslhangindent}
\newlength{\csllabelwidth}
 {\begin{list}{}{%
  \setlength{\itemindent}{0pt}
  \setlength{\leftmargin}{0pt}
  \setlength{\parsep}{0pt}
  \ifodd #1
   \setlength{\leftmargin}{\cslhangindent}
   \setlength{\itemindent}{-1\cslhangindent}
  \fi
  \setlength{\itemsep}{#2\baselineskip}}}
 {\end{list}}
\usepackage{calc}

\usepackage{amsmath} \usepackage{array} \usepackage{float}

\begin{document}

\sectionhead{Contributed research article}
\volume{XX}
\volnumber{YY}
\year{20ZZ}
\month{AAAA}

\begin{article}
\title{cardinalR: Generating Interesting High-Dimensional Data Structures}

\author{by Jayani P. Gamage, Dianne Cook, Paul Harrison, Michael Lydeamore, and Thiyanga S. Talagala}

\maketitle

\abstract{%
Simulated high dimensional data is useful for testing, validating, and improving algorithms used in dimension reduction, supervised and unsupervised learning. High-dimensional data is characterized by multiple variables that are dependent or associated in some way, such as linear, nonlinear, clustering or anomalies. Here we provide new methods for generating a variety of high-dimensional structures using mathematical functions and statistical distributions organized into the R package cardinalR. Several example data sets are also provided. These will be useful for researchers to better understand how different analytical methods work and can be improved, with a special focus on nonlinear dimension reduction methods. This package enriches the existing toolset of benchmark datasets for evaluating algorithms.
}

\section{Introduction}\label{introduction}

Generating synthetic datasets with clearly defined geometric properties is useful for evaluating and benchmarking algorithms in various fields, such as machine learning, data mining, and computational biology. Researchers often need to generate data with specific dimensions, noise characteristics, and complex underlying structures to test the performance and robustness of their methods. There are numerous packages available in R for generating synthetic data, each designed with unique characteristics and focus areas. The \texttt{geozoo} package \citep{barret2016} provides functions to generate standard high-dimensional data like cubes, spheres and simplexes, along with some prepared datasets. The \texttt{snedata} package \citep{james2025} provides functions for generating common examples used in dimension reduction publications and to download benchmark data sets. The \texttt{splatter} package \citep{luke2017} is designed to simulate complex biological data, capturing field-specific nuances such as batch effects and differential expression. The \texttt{mlbench} package \citep{friedrich2024} provides access to benchmark datasets commonly associated with established classification or regression challenges. The \texttt{surreal} package \citep{james2024} implements the ``Residual (Sur)Realism'' algorithm \citep{leonard2007} to generate datasets that embed hidden images or text into residual plots, providing engaging visual demonstrations for teaching model diagnostics.

The current work implemented in the \texttt{cardinalR} R package builds on these approaches. It provides functions to generate a more extensive set of high-dimensional data structures, allowing users to: (i) construct high-dimensional datasets based on geometric shapes, including the option to enhance dimensionality by adding controlled noise dimensions; (ii) introduce adjustable levels of background noise to these structures; and (iii) combine the shapes to produce multiple clusters. The user can control characteristics such as number of dimensions, shape and sample size. It is designed to resource researchers with synthetic datasets to evaluate the performance and interpret the fit of NLDR methods, clustering algorithms, and visualization techniques. These datasets can also serve as benchmark examples for exploring how different choices of algorithm parameters affect the identification or representation of cluster and manifold structures in high-dimensional spaces.

The motivation for developing this package originated from our own work in studying nonlinear dimension reduction (NLDR) algorithms. We wanted to conduct a visualization experiment to understand perception and misperception of a variety of NLDR methods. This required simulated datasets with carefully controlled geometric and clustering properties. While some existing packages provided useful starting points, none fully supported the creation of flexible, high-dimensional data with the specific structural variations needed for our experiment. Developing these generators for research purposes underlies \texttt{cardinalR}, which is now a general-purpose package that should be useful for research and teaching.

The example data structures are best viewed using a tour \citep{As85}. These show the data as a sequence of low dimensional projections (typically \(2\text{-}D\)), providing a good sense of the shape in high dimensions. The interactive tour plots included in this paper are produced using the software \texttt{langevitour} \citep{paul2023}.

The next section provides an overview of the usage of the \texttt{cardinalR} package, illustrating how its modular components can be combined to generate complex high-dimensional datasets. This is followed by a section describing the implementation of the package, including its design principles and key functions. The Application section then demonstrates how the simulated clustering structures can be used to evaluate and compare dimension reduction and clustering methods. Finally, we give a brief conclusion of the paper and discuss potential opportunities for the use of our data collection.

\section{Usage}\label{usage}

The \texttt{cardinalR} package is built on a modular framework where individual geometric generators (e.g., Gaussian, cone, sphere) create well-defined shapes (A full list of available shape generators are available at \url{https://jayanilakshika.github.io/cardinalR/reference/index.html}.), which can then be combined into a single dataset including scaling, rotation, and translation. The package is available on CRAN, and the source is available on GitHub at \href{https://github.com/JayaniLakshika/cardinalR}{JayaniLakshika/cardinalR}.

\begin{figure}[!ht]

{\centering \includegraphics[width=0.8\linewidth,alt={The figure is a workflow diagram showing the process for generating high-dimensional clustered data. Boxes or nodes represent steps such as specifying input parameters (number of points, number of clusters, cluster locations, shapes, scaling, rotations, and optional background noise), generating each cluster with a shape generator, optionally rotating or scaling clusters, combining them into a single dataset, adding background noise if desired, and labeling each observation by cluster shape. Arrows indicate the flow of the process from parameter specification to the final labeled dataset. The diagram uses a clear layout with distinct steps, showing how data are constructed in sequence.}]{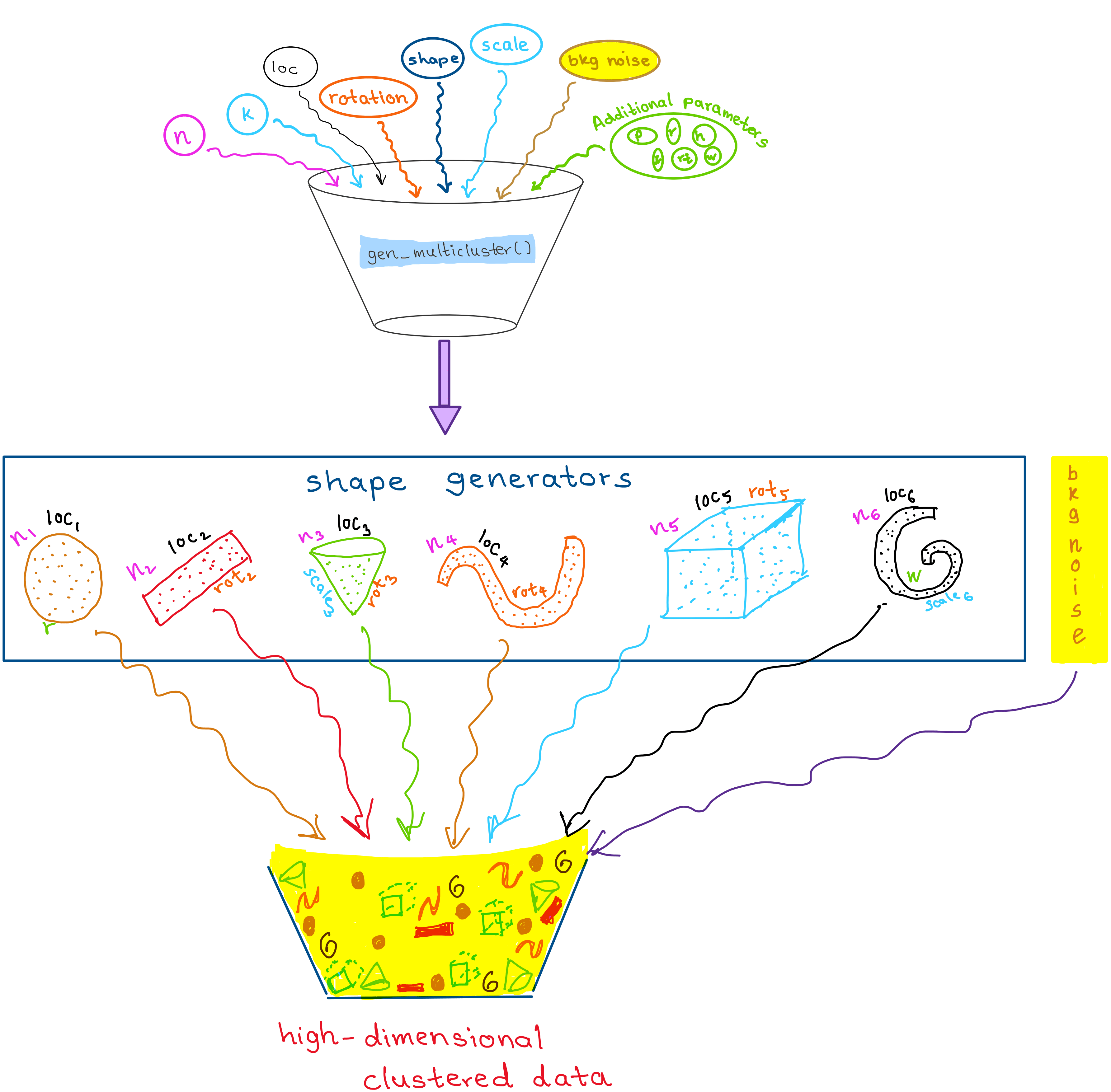} 

}

\caption{Workflow for generating high-dimensional clustered data. The user specifies input parameters such as the number of points ($n$), number of clusters ($k$), cluster locations, shapes, scaling, rotations, and optional background noise. Each cluster shape is generated by a shape generator, optionally rotated or scaled, and combined into a single dataset. Additional background noise can be added, and each observation is labeled by shape.}\label{fig:workflow}
\end{figure}

The main function, \texttt{gen\_multicluster()}, is an all-in-one function that includes generating individual shapes, handling scaling and rotating of these shapes, and combining the result into a single unified dataset. This function and associated workflow allow flexible construction of complex, high-dimensional structures for evaluating clustering and dimension reduction methods. Figure \ref{fig:workflow} illustrates the workflow of \texttt{gen\_multicluster()}.

Users can control the number of clusters (\texttt{k}), and the number of points in each cluster (\texttt{n}). Each cluster can take on a different geometric shape (e.g., Gaussian, cone, uniform cube) by specifying the corresponding generator function (\texttt{shape}), can be scaled to adjust its spread, rotated using custom rotation matrices (\texttt{rotation}), and positioned at defined centroids (\texttt{loc}). The function ensures flexibility in cluster location and orientation, allowing users to simulate complex high-dimensional structures.

The following is an example of a three-shape multiclustered dataset. The first shape is Gaussian, the second conical, and the third a cube.

\begin{verbatim}
clust_data <- gen_multicluster(
  n = c(200, 300, 500),
  k = 3,
  loc = matrix(c(
    0, 0, 0, 0,
    5, 9, 0, 0,
    3, 4, 10, 7
  ), nrow = 3, byrow = TRUE),
  scale = c(3, 1, 2),
  shape = c("gaussian", "cone", "unifcube"),
  is_bkg = FALSE
)
\end{verbatim}

Here, the shapes have \(200\), \(300\), and \(500\) points respectively (\texttt{n}), are positioned in \(4\text{-}D\) space according to a location matrix, \texttt{loc}, and stretched according to the \texttt{scale}. The details of the individual shape generators and the noise elements are contained in the following sections.

\section{Implementation}\label{implementation}

The main function of the package is \texttt{gen\_multicluster()}, which generates datasets consisting of multiple clusters with user-specified characteristics.
To maintain consistency across generators, the function identifies the arguments required by each chosen generator function and supplies only those arguments that are valid for that specific generator. This design enables the combination of cluster types with differing parameter requirements within the same dataset. When clusters are generated with fewer dimensions than others, the function augments the lower-dimensional clusters with additional Gaussian noise variables so that all clusters are represented in the same dimensional space. These noise dimensions are drawn independently from normal distributions \(X \sim \mathcal{N}(\mu, \sigma^2)\), where the mean (\(\mu\)) is set to the average of the cluster coordinates and the standard deviation (\(\sigma\)) defaults to \(0.2\).

An optional argument, \texttt{is\_bkg}, adds background noise drawn from a multivariate normal distribution centered on the dataset's overall mean with standard deviations matching the observed spread. Extra arguments (\texttt{...}) can be passed to cluster generators, allowing further control over per-cluster characteristics like radius of the sphere. The main arguments of the \texttt{gen\_multicluster()} function are shown in Table \ref{tab:main-tb-pdf}.

\begin{table}

\caption{\label{tab:main-tb-pdf}The main arguments for \texttt{gen\_multicluster()}.}
\centering
\begin{tabular}[t]{>{\raggedright\arraybackslash}p{2cm}>{\raggedright\arraybackslash}p{3cm}>{\raggedright\arraybackslash}p{8cm}}
\toprule
Argument & Type & Explanation\\
\midrule
\texttt{n} & integer (vector) & Number of points in each cluster.\\
\texttt{k} & integer & Number of clusters.\\
\texttt{loc} & numeric (matrix) & Locations/centroids of clusters.\\
\texttt{scale} & numeric (vector) & Scaling factors of clusters.\\
\texttt{shape} & character (vector) & Shapes of clusters.\\
\texttt{rotation} & numeric (list) & Rotation matrices, one per cluster.\\
\texttt{is\_bkg} & boolean & Background noise should exist or not.\\
\bottomrule
\end{tabular}
\end{table}

\subsection{Shape generators}\label{shape-generators}

The shape generators form the foundation of the package, providing functions to create synthetic datasets from simple, well-defined geometric forms such as cones, pyramids, spheres, grids, and branching structures. Each generator includes the parameter \texttt{n}, which specifies the number of points to generate. Some functions, such as \texttt{gen\_unifcube()}, also take the dimension \texttt{p}, while others include arguments specific to the geometry (e.g., radius for spheres (\texttt{r}), width for bands (\texttt{w})). If higher-dimensional data are required, additional noise dimensions can be appended after data generation using any noise generator function. This flexibility allows users to construct both low- and high-dimensional datasets from the same underlying structures. Table \ref{tab:shape-tb-pdf} outlines these functions. The main arguments of the functions described in Table \ref{tab:arg-shape-tb-pdf}.

\begin{table}

\caption{\label{tab:shape-tb-pdf}Overview of shape-generation functions, including their required parameters and a brief description of each geometric structure produced. The generators cover branching patterns, spheres, spirals, pyramids, Gaussian clouds, and other nonlinear shapes.}
\centering
\begin{tabular}[t]{>{\raggedright\arraybackslash}p{3.5cm}>{\raggedright\arraybackslash}p{1.5cm}>{\raggedright\arraybackslash}p{8cm}}
\toprule
Function & Arguments & Explanation\\
\midrule
\texttt{gen\_expbranches} & \texttt{n, k} & Exponential shaped branches.\\
\texttt{gen\_linearbranches} & \texttt{n, k} & Linear shaped branches.\\
\texttt{gen\_curvybranches} & \texttt{n, k} & Curvy shaped branches.\\
\texttt{gen\_orglinearbranches} & \texttt{n, p, k} & Linear shaped branches originated in one point.\\
\texttt{gen\_orgcurvybranches} & \texttt{n, p, k} & Curvy shaped branches originated in one point.\\
\texttt{gen\_cone} & \texttt{n, p, h, ratio} & Cone-shaped structure.\\
\texttt{gen\_gridcube} & \texttt{n, p} & Cube with specified grid points along each axes.\\
\texttt{gen\_unifcube} & \texttt{n, p} & Cube with uniform points.\\
\texttt{gen\_gaussian} & \texttt{n, p, s} & Multivariate Gaussian cloud.\\
\texttt{gen\_longlinear} & \texttt{n, p} & Long linear structure.\\
\texttt{gen\_mobius} & \texttt{n} & Möbius strip in $3\text{-}D$.\\
\texttt{gen\_quadratic} & \texttt{n} & Quadratic pattern in $2\text{-}D$.\\
\texttt{gen\_cubic} & \texttt{n} & Cubic pattern in $2\text{-}D$.\\
\texttt{gen\_pyrrect} & \texttt{n, p, l\_vec, rt} & Rectangular-base Pyramid, with a sharp or blunted apex.\\
\texttt{gen\_pyrtri} & \texttt{n, p, h, l, rt} & Triangular-base Pyramid, with a sharp or blunted apex.\\
\texttt{gen\_pyrstar} & \texttt{n, p, h, rb} & Star-shaped base Pyramid, with a sharp or blunted apex.\\
\texttt{gen\_pyrfrac} & \texttt{n, p} & Pyramid with triangular pyramid-shaped holes.\\
\texttt{gen\_scurve} & \texttt{n} & S-curve in $3\text{-}D$.\\
\texttt{gen\_circle} & \texttt{n, p} & Circle.\\
\texttt{gen\_curvycycle} & \texttt{n, p} & Curvy cell cycle.\\
\texttt{gen\_unifsphere} & \texttt{n, r} & Uniform ball.\\
\texttt{gen\_hollowsphere} & \texttt{n, p} & Hollow sphere.\\
\texttt{gen\_gridedsphere} & \texttt{n} & Grided sphere.\\
\texttt{gen\_clusteredspheres} & \texttt{n, k, r, loc} & Multiple small spheres within a big sphere.\\
\texttt{gen\_hemisphere} & \texttt{n, p} & Hemisphere.\\
\texttt{gen\_swissroll} & \texttt{n, w} & Swissroll structure.\\
\texttt{gen\_trefoil4d} & \texttt{n, steps} & Trefoil in $4\text{-}D$.\\
\texttt{gen\_trefoil3d} & \texttt{n, steps} & Trefoil in $3\text{-}D$.\\
\texttt{gen\_crescent} & \texttt{n} & Crescent pattern.\\
\texttt{gen\_curvycylinder} & \texttt{n, h} & Curvy cylinder.\\
\texttt{gen\_sphericalspiral} & \texttt{n, spins} & Spherical spiral.\\
\texttt{gen\_helicalspiral} & \texttt{n} & Helical spiral.\\
\texttt{gen\_conicspiral} & \texttt{n, spins} & Conic spiral.\\
\texttt{gen\_nonlinear} & \texttt{n, hc, non\_fac} & Nonlinear hyperbola.\\
\bottomrule
\end{tabular}
\end{table}

\begin{table}

\caption{\label{tab:arg-shape-tb-pdf}Argument definitions for the shape generators. The table lists each argument, its data type, and a description of its role in controlling geometric structure, dimensionality, scaling, curvature, spacing, and other features of the simulated high-dimensional datasets.}
\centering
\begin{tabular}[t]{>{\raggedright\arraybackslash}p{2cm}>{\raggedright\arraybackslash}p{3cm}l}
\toprule
Argument & Type (positive) & Explanation\\
\midrule
\texttt{n} & integer & Number of points.\\
\texttt{k} & integer & Number of clusters.\\
\texttt{p} & integer & Number of dimensions.\\
\texttt{h} & real value & Height.\\
\texttt{ratio} & real value & Radius tip to radius base ratio.\\
\addlinespace
\texttt{s} & real value & Variance-covariance matrix.\\
\texttt{r} & real value & Radius.\\
\texttt{n\_vec} & integers & Sample sizes of the big and small spheres\\
\texttt{k\_small} & integer & Number of small spheres.\\
\texttt{r\_vec} & real values & Radius of the big and small spheres.\\
\addlinespace
\texttt{spe} & real value & How far apart the small spheres are placed.\\
\texttt{w} & real value & Vertical variation\\
\texttt{steps} & integer & Number of steps for the theta parameter.\\
\texttt{spins} & integer & Number of loops of the spiral.\\
\texttt{hc} & real value & Steepness and vertical scaling of the hyperbola.\\
\addlinespace
\texttt{non\_fac} & real value & Strength of this sinusoidal effect.\\
\texttt{l} & real value & Base length of the pyramid.\\
\texttt{l\_vec} & real values & Base lengths along the and y of the pyramid.\\
\texttt{rt} & real value & Tip radius of the pyramid.\\
\texttt{rb} & real value & Base radius of the pyramid.\\
\bottomrule
\end{tabular}
\end{table}

\subsubsection{Branching}\label{branching}

\begin{figure}[!ht]
\includegraphics[width=1\linewidth,alt={The figure consists of three 2‑D scatter plots, each showing a different 2‑D projection from the 4‑D orgcurvybranches dataset. In each plot, the horizontal axis represents one coordinate of the selected 2‑D projection, and the vertical axis represents the other coordinate. Points are plotted as individual markers and appear grouped into multiple elongated clusters or branches, which curve or radiate through the plot. Each branch is distinguished by color, with overlapping branches occurring when subspaces are shared. The plots have roughly square aspect ratios, and together they show how the same high‑dimensional branching structure appears differently from three viewing angles.}]{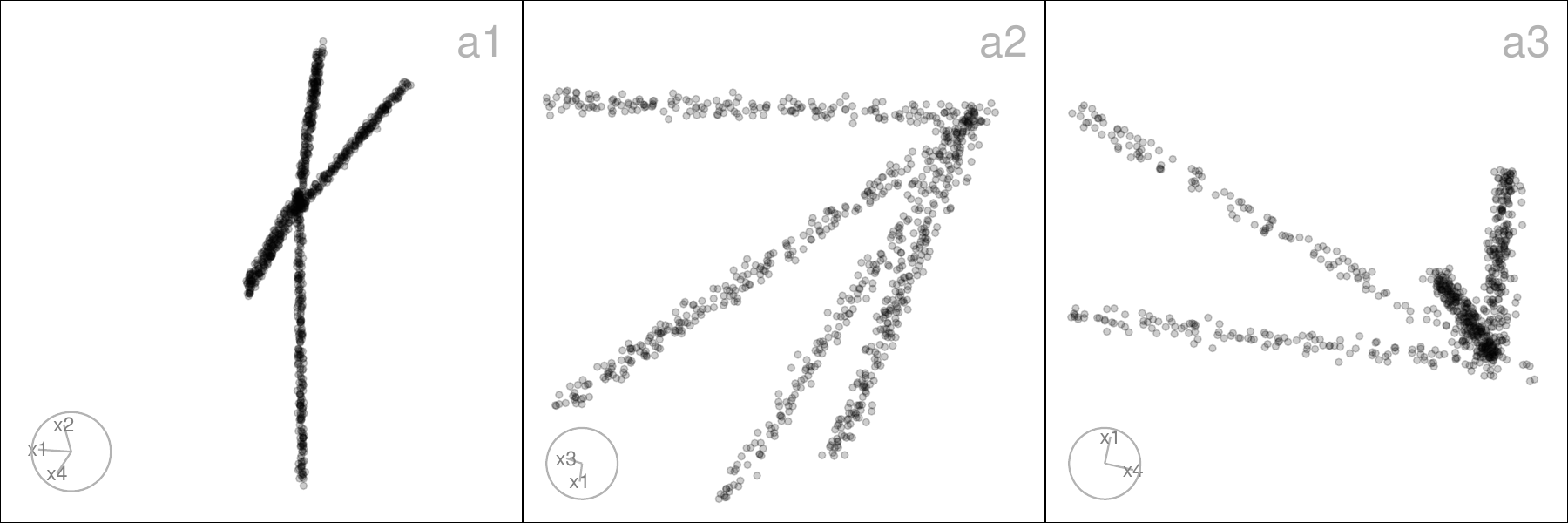} \caption{Three $2\text{-}D$ projections from the $4\text{-}D$ \texttt{orgcurvybranches} data. Each shows a different projection, illustrating how the linear branches appear from multiple viewing angles. These views highlight the dataset’s underlying branching structure and demonstrate how projections reveal patterns that are otherwise hidden in higher dimensions.}\label{fig:branch-proj}
\end{figure}

A branching structure (Figure \ref{fig:branch-proj}) captures trajectories that diverge or bifurcate from a common origin, similar to processes such as cell differentiation in biology \citep{trapnell2014}. We introduce a set of data generation functions specifically designed to simulate high-dimensional branching structures with various geometries, total number of points (\texttt{n}) generated across all branches, with points allocated approximately evenly among branches, and number of branches (\texttt{k}). Although these functions can generate multiple branches, they do not produce a formal \emph{multicluster} dataset: the branches form a single connected structure, with multiple visually distinct arms rather than independent clusters.

The simplest structures are approximately linear branches in \(2\text{-}D\), generated by the \texttt{gen\_linearbranches(n,\ k)} function. These consist of \(k\) short line segments in the first two dimensions, with added jitter to simulate variability. Mathematically, each branch \(i\) is defined as

\[
X_1 \sim U(a_i, b_i), \quad X_2 = s_i (X_1 - x_{\text{start},i}) + y_{\text{start},i} + \epsilon, \quad \epsilon \sim U(0, \delta),
\]

where \((x_{\text{start},i}, y_{\text{start},i})\) is the starting point of branch \(i\), \(\delta\) controls local jitter, and \(s_i\) is the slope, initialized as

\[
s_i =
\begin{cases}
0.5 & i = 1, \\
-0.5 & i = 2, \\
\text{randomly sampled from } [s_{\min}, s_{\max}] & i = 3, \dots, k.
\end{cases}
\]

The jitter term is sampled from a one-sided uniform distribution to introduce directional variability without altering branch orientation.

Branches \(1\) and \(2\) are initialized with fixed slopes and starting points, while later branches are iteratively added at locations chosen to avoid overlap with existing branches, producing a set of connected linear paths.

To introduce curvature, the \texttt{gen\_curvybranches(n,\ k)} function generates \(k\) curvilinear branches in \(2\text{-}D\). Each branch follows a quadratic trajectory of the form

\[
X_1 \sim U(a_i, b_i), \quad X_2 = 0.1 X_1 + s_i X_1^2 + \epsilon, \quad \epsilon \sim U(-\delta, \delta),
\]
where \((a_i, b_i)\) defines the domain of the branch, \(s_i\) controls curvature, and \(\delta\) introduces local jitter. For the first two branches, the parameters are fixed to establish reference shapes: \((a_1,b_1,s_1) = (0,1,1), \quad (a_2,b_2,s_2) = (-1,0,-2)\). Additional branches are attached iteratively to existing structures. Each new branch \(i\) starts at a selected point \((x_{\text{start},i}, y_{\text{start},i})\) from the current structure and extends according to

\[
X_1 \sim U(x_{\text{start},i}, x_{\text{start},i}+1), \quad X_2 = 0.1 X_1 - s_i (X_1^2 - x_{\text{start},i}) + y_{\text{start},i},
\]

where \(s_i\) is a scale factor controlling the curvature of branch \(i\). For the first few initial branches, \(s_i\) can be fixed (e.g., \(s_1 = 1, s_2 = 2\)) to establish reference shapes, while for subsequent branches it is sampled from a predefined set, such as \(s_i \in \{-2, -1.5, -1, -0.5, 0, 0.5, 1, 1.5\}\), to control curvature magnitude and direction.

The \texttt{gen\_expbranches(n,\ k)} function creates \(k\) exponential branches in \(2\text{-}D\), radiating from a central region. Each branch \(i\) is defined as

\[
X_1 \sim U(-2,2), \quad X_2 = \exp(\sigma_i \, s_i \, X_1) + \epsilon, \quad \epsilon \sim U(0, \delta), \quad s_i \sim U(0.5,2),
\]

where \(\sigma_i = (-1)^{i+1}\) alternates the sign of the exponent to produce mirror-symmetric branches. The parameter \(s_i\) controls the steepness of branch \(i\), and \(\delta\) introduces small local jitter.

High-dimensional generalizations are provided by \texttt{gen\_orglinearbranches(n,\ p,\ k)} (Figure \ref{fig:branch-proj}) and \texttt{gen\_orgcurvybranches(n,\ p,\ k)}. For branch
\(i\), the active coordinate pair \((i_1, i_2)\) indexes the selected \(2\text{-}D\) subspace. When \texttt{allow\_share\ =\ TRUE}, multiple branches may share the same subspace; otherwise, subspaces are sampled without replacement until all possible \(\binom{p}{2}\) combinations are exhausted, after which additional branches may repeat subspaces.

In both cases, branch \(i\) is generated according to

\[
X_{i_1} \sim U(a_i, b_i), \quad
X_{i_2} = f_i(X_{i_1}) + \epsilon, \quad
\epsilon \sim N(0, \sigma^2),
\]
where \(a_i\) and \(b_i\) define the domain of the branch and \(\epsilon\) introduces smooth variability in the \(p\text{-}D\) space. The function \(f_i(\cdot)\) determines the branch geometry:
\[
f_i(x) =
\begin{cases}
s_i x, & \text{linear branches}, \\
-s_i x^2, & \text{curvilinear branches}.
\end{cases}
\]

The scale factor \(s_i\) controls slope (linear branches) or curvature (curvilinear branches) and is assigned as follows: for the first \(\binom{p}{2}\) branches, \(s_i = 1\); for additional branches when \(k > \binom{p}{2}\), \(s_i\) is randomly drawn from the set \(\{1, 1.5, 2, \dots, 8\}\).

Across all branching generators, the scale parameter \(s_i\) controls the strength of deviation from linearity, determining slope, curvature, or growth rate depending on the branch geometry.

\subsubsection{Cone}\label{cone}

\begin{figure}[!ht]
\includegraphics[width=1\linewidth,alt={The figure consists of three 2‑D scatter plots, each showing a different 2‑D projection of the 4‑D cone dataset. In each plot, the horizontal and vertical axes represent two continuous numeric dimensions. Points are plotted individually and are concentrated near one end, forming a narrow tip, and gradually spread out along the other direction, creating a wider base. The points trace an elongated, triangular or cone-like shape in each projection. All points use the same color and marker, with no additional visual encodings, and the plots have roughly square aspect ratios. Together, the three projections illustrate the distribution of points along the conical structure from different viewing angles.}]{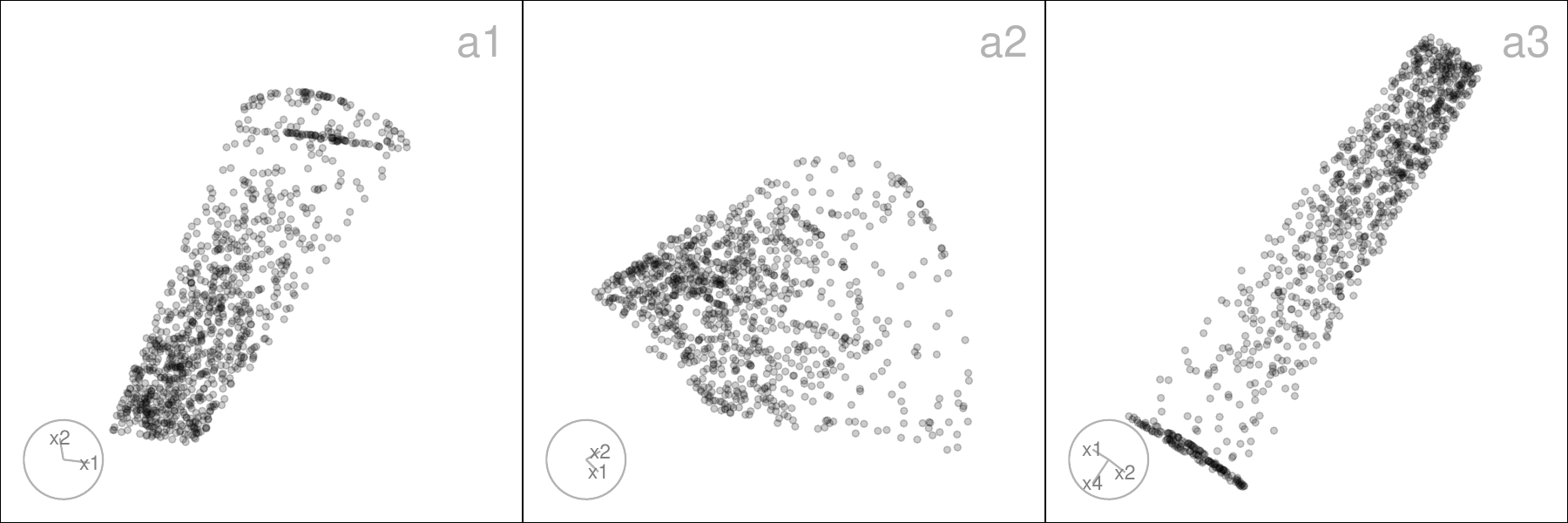} \caption{Three $2\text{-}D$ projections from the $4\text{-}D$ \texttt{cone} data. Points are concentrated near the tip along the height dimension, while the radius of the hyperspherical cross-section decreases linearly toward the apex. These projections show how the conical geometry is preserved.}\label{fig:cone-proj}
\end{figure}

To simulate a cone-shaped structure in arbitrary dimensions (Figure \ref{fig:cone-proj}), we define a function \texttt{gen\_cone(n,\ p,\ h,\ ratio)}, which creates a high-dimensional cone with options for a sharp or blunted apex, allowing for a dense concentration of points near the tip.

This function generates \(n\) points in \(p\text{-}D\), where the last dimension, \(X_p\), represents the height along the cone's axis, and the first \(p-1\) dimensions define a shrinking hyperspherical cross-section toward the tip. Heights are sampled from a truncated exponential distribution, \(X_p \sim \text{Exp}(\lambda = 2/h)\), truncated to the interval \([0, h]\), producing a higher density of points near the tip. At each height \(X_p\), the radius of the cross-section increases linearly from base to tip according to \(r = r_{\text{min}} + (r_{\text{max}} - r_{\text{min}}) X_p / h\), where \(r_{\text{min}} = \text{ratio} \in [0, 1]\) and \(r_{\text{max}} = 1\).

For each point, a direction in the first \(p-1\) dimensions is sampled uniformly on a \((p-1)\)-dimensional hypersphere using generalized spherical coordinates. The radial coordinates are scaled by the height-dependent radius \(r\), producing the conical taper. In three dimensions (\(p = 3\)), this results in a classical \(3\text{-}D\) cone, while for \(p > 3\), additional dimensions provide a smooth embedding into higher-dimensional space, preserving the conical structure.

Cone-shaped structures appear in particle dispersions, light beams, and tapering processes, where spread decreases along one axis. They are also used to benchmark clustering and dimensionality reduction methods \citep{hadsell2006}.

\subsubsection{Cube}\label{cube}

A cube structure represents uniformly or systematically distributed points within a high-dimensional hypercube, providing a useful framework for assessing how well algorithms preserve uniformity, and boundary properties in high dimensions. We provide a set of functions to generate high-dimensional cube structures with flexible configurations, including regular grids, and uniform random points.

The function \texttt{gen\_gridcube(n,\ p)} is a wrapper around \texttt{geozoo::cube.solid.grid()}. It generates a regular lattice of points in \(p\text{-}D\), producing a uniform hypercube grid. The parameter \texttt{n} controls the approximate number of points by determining the grid resolution along each axis.

By contrast, \texttt{gen\_unifcube(n,\ p)} wraps \texttt{geozoo::cube.solid.random()}, producing uniformly distributed points within a \(p\text{-}D\) cube. To avoid including the cube's vertices, these points are removed after generation. This results in a hypercube filled with random samples rather than structured lattice points.

Such cube-based structures are commonly used as benchmarks in Monte Carlo sampling, computational geometry, and density estimation, where assessing how algorithms behave under uniform or grid-like distributions is critical \citep{devroye1986, niederreiter1992}.

\subsubsection{Gaussian}\label{gaussian}

The \texttt{gen\_gaussian(n,\ p,\ s)} function generates a multivariate Gaussian cloud in \(p\text{-}D\), centered at the origin with user-defined covariance structure. For
\(i=1,\dots,n\), each observation is independently drawn from a multivariate normal distribution, \(X_i \sim N_p(\boldsymbol{0}, s)\), where \(s\) is a user-defined \(p \times p\) positive-definite matrix.

Gaussian clouds are common benchmark structures in statistics and machine learning, used in clustering, classification, and anomaly detection, with applications in image segmentation, speech recognition, and forensic analysis \citep{geoffrey2000}.

\subsubsection{Linear}\label{linear}

The \texttt{gen\_longlinear(n,\ p)} function generates a high-dimensional dataset representing a single noisy linear trajectory. Let \(t_i = i - 1, \quad i = 1, \dots, n\), denote a common latent index shared across all dimensions. For each dimension \(j = 1, \dots, p)\), independent scale and shift parameters are sampled as \(a_j \sim U(-10, 10), \qquad b_j \sim U(-300, 300)\). Gaussian noise \(\varepsilon_{ij} \sim N(0, (0.03n)^2)\) is added independently across observations and dimensions. The observed variables are then defined as \(X_{ij} = a_j \bigl(t_i + b_j + \varepsilon_{ij}\bigr), \quad i = 1, \dots, n\). This construction yields a single elongated linear structure embedded in \(p\text{-}D\), with each dimension exhibiting a different orientation, scale, and offset.

This structure appears in \(p\text{-}D\) data when variation is driven by a single factor, such as time-course or sensor measurements, providing a useful test case for trajectory and regression methods \citep{trapnell2014}.

\subsubsection{Möbius}\label{muxf6bius}

The \texttt{gen\_mobius()} function is a wrapper around \texttt{geozoo::mobius()}, designed to simplify the generation of a Möbius strip in three dimensions for use in high-dimensional diagnostic studies. The function returns a tibble with \(n\) sampled points forming the surface of a Möbius strip.

The Möbius strip structure can model twisted or cyclic surfaces in physics and engineering, such as conveyor belts, molecular structures, or optical systems with non-orientable geometries \citep{optica2023}.

\subsubsection{Polynomial}\label{polynomial}

A polynomial structure generates data points that follow nonlinear curvilinear relationships, such as quadratic or cubic trends, in \(2\text{-}D\) space. To extend these patterns into high-dimensional settings, additional noise dimensions can be added. These patterns are useful for evaluating how well algorithms capture smooth, nonlinear trajectories and curvature in the data. We provide functions for generating quadratic and cubic structures, enabling controlled experiments with different degrees of polynomial complexity.

The first is the quadratic curve of \(n\) points in two dimensions. This is generated using \texttt{gen\_quadratic(n,\ range)}. Let \(range = [a, b]\). The independent variable is defined as \(X_1 \sim U(a, b)\), and the response is generated as \(X_2 = X_1 - X_1^2 + \varepsilon\), where \(\varepsilon \sim U(0, 0.5)\). This produces a smooth parabolic arc opening downward, with vertical jitter introduced by the noise term.

The second is the cubic curve of \(n\) points in two dimensions. This is generated using \texttt{gen\_cubic(n,\ range)}. Let \(range = [a, b]\). The independent variable is defined as \(X_1 \sim U(a, b)\), and a raw polynomial basis of degree \(3\) is applied to construct \(X_2 = X_1 + X_1^2 - X_1^3 + \varepsilon_2\), where \(\varepsilon_2 \sim U(0, 0.5)\). This produces a more complex curvilinear structure than the quadratic case, with both upward and downward turning points.

\subsubsection{Pyramid}\label{pyramid}

\begin{figure}[!ht]

{\centering \includegraphics[width=0.8\linewidth,alt={The figure shows three 2‑D scatter plots for each of four 4‑D datasets: pyrrrect, pyrtri, pyrstar, and pyrholes. In each plot, the horizontal and vertical axes represent continuous numeric coordinates of the selected 2‑D projection, and points are plotted individually as small, uniformly colored markers with no additional visual encodings. In pyrrrect, points form a dense rectangular base tapering to a narrow tip; in pyrtri, points form a triangular distribution with sharper edges; in pyrstar, points extend into multiple pointed branches radiating from a central region; and in pyrholes, points cluster in a compact shape but leave hollow or open regions within it. The plots maintain roughly square aspect ratios and illustrate how these different pyramid-like structures appear when projected from four to two dimensions.}]{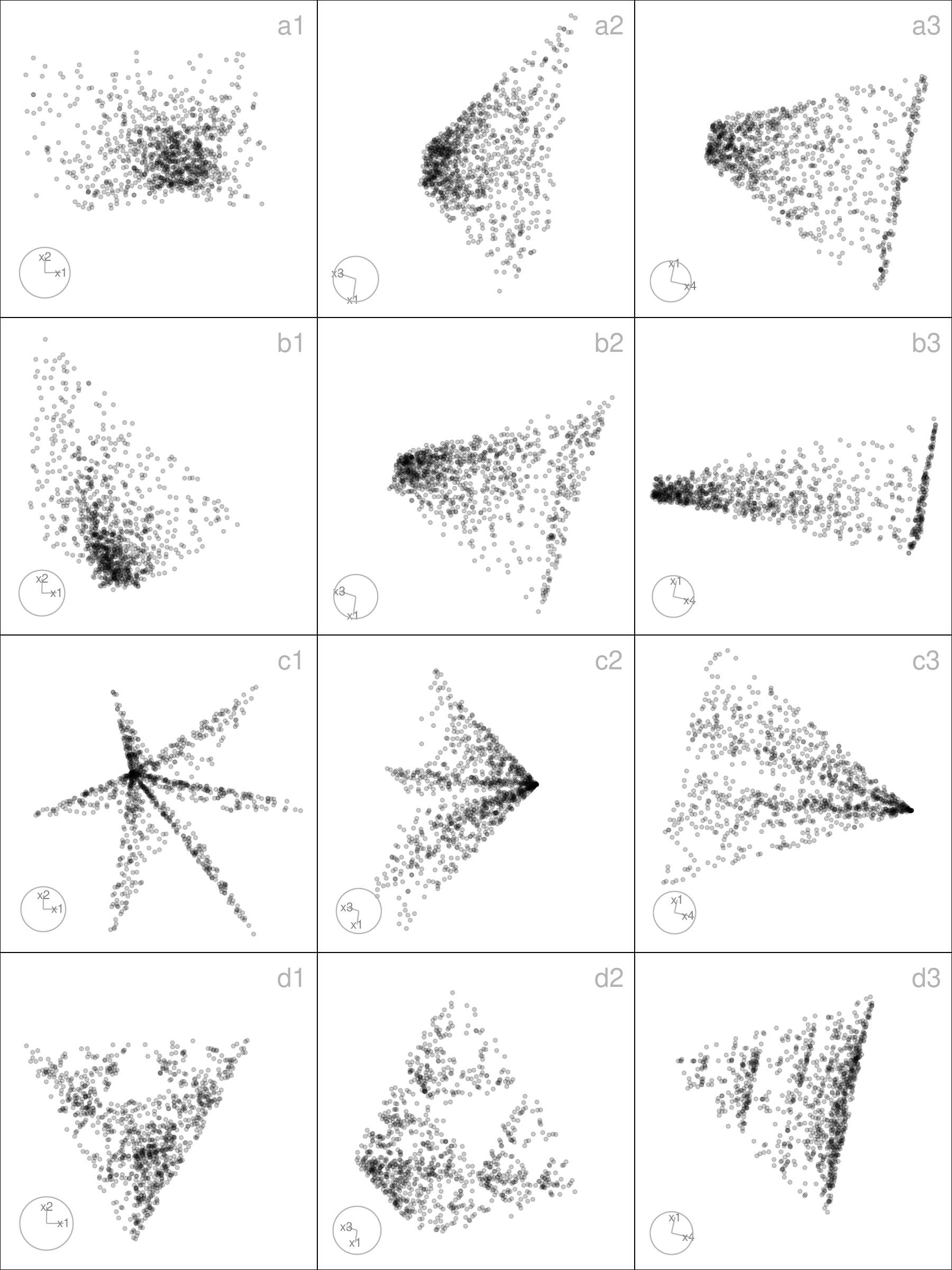} 

}

\caption{Three $2\text{-}D$ projections from $4\text{-}D$, for the \texttt{pyrrect} (a1-a3), \texttt{pyrtri} (b1-b3), \texttt{pyrstar} (c1-c3), and \texttt{pyrholes} (d1-d3) data. The \texttt{pyrrrect} structure forms a dense rectangular base tapering to a narrow tip, while \texttt{pytri} shows a more triangular spread with sharper edges. \texttt{pyrstar} extends into multiple pointed branches radiating from a common core, and \texttt{pyrholes} reveals hollow or open regions within an otherwise compact shape. These projections illustrate a range of pyramid-like geometries that vary in density and structure.}\label{fig:pyr-proj}
\end{figure}

A pyramid structure (Figure \ref{fig:pyr-proj}) represents data arranged around a central apex and base, useful for exploring how algorithms handle pointed or layered geometries in \(p\text{-}D\) space. The functions provided allow users to generate pyramids with rectangular, triangular, and star-shaped bases, and sharp or blunted apexes. Additionally, it is possible to create a pyramid with a fractal-like internal structure, enabling the study of non-convex and sparse regions.

Let \(X_1, \dots, X_p\) denote the coordinates of the generated points. For the rectangular, triangular, and star-shaped based pyramid generator functions, the final dimension, \(X_p\), encodes the height of each point and is drawn from an exponential distribution capped at the maximum height \(h\). That is, \(X_p = z \sim \min\left(\text{Exp}(\lambda = 2/h),\ h\right).\) This distribution creates a natural skew toward smaller height values, resulting in a denser concentration of points near the pyramid's apex. For the star-shaped base pyramid, the final dimension is drawn from a uniform distribution. That is, \(X_p = z \sim U(0, h)\).

The remaining dimensions are based on the specific pyramid shape. For the rectangular based pyramid, \texttt{gen\_pyrrect(n,\ p,\ h,\ l\_vec,\ rt)} (Figure \ref{fig:pyr-proj} a) the base shape is a rectangle whose size shrinks linearly with height. Let \(l_x\) and \(l_y\) denote the half-widths of the rectangular base in the \(X_1\) and \(X_2\) directions, specified via \(l_{vec}=(l_x,l_y)\), and let \(r_t\) denote the half-width at the pyramid tip. At height \(z\in [0,h]\), the half-widths of the rectangular cross-section are \(r_x(z) = r_t + (l_x - r_t)z/h\), \(r_y(z) = r_t + (l_y - r_t)z/h\). The first three coordinates are then defined as \(X_1 \sim U(-r_x(z),\ r_x(z)), \quad X_2 \sim U(-r_y(z),\ r_y(z)),\text{ and }X_3 \sim U(-r_x(z),\ r_x(z))\).

For the triangular based pyramid, \texttt{gen\_pyrtri(n,\ p,\ h,\ l,\ rt)} (Figure \ref{fig:pyr-proj} b), let \(r(z)\) denote the scaling factor (distance from the origin to triangle vertices) at height \(z\). That is, \(r(z) = r_t + (l-r_t)z/h\). A point in the triangle at height \(z\) is generated using barycentric coordinates \((u, v)\) to ensure uniform sampling within the triangular cross-section: \(u, v \sim U(0, 1), \quad \text{if } u + v > 1: u \leftarrow 1 - u,\ v \leftarrow 1 - v\). The first three coordinates (triangle plane) are then: \(X_1 = r(z)(1 - u - v)\), \(X_2 = r(z)u\), and \(X_3 = r(z)v.\)

For the star based pyramid, \texttt{gen\_pyrstar(n,\ p,\ h,\ rb)} (Figure \ref{fig:pyr-proj} c), let the radius at height \(z\), \(r(z)\), be such that the radius scales linearly from zero (tip) to the base radius \(r_b\). That is, \(r(z) = r_b\left(1 - z/h\right)\). Each point is placed within a regular hexagon in the plane \((X_1, X_2)\), using a randomly chosen hexagon sector angle \(\theta \in \{0, \pi/3, 2\pi/3, \pi, 4\pi/3, 5\pi/3\}\) and a uniformly random radial scaling factor: \(\theta \sim DiscreteUniform\{0, \pi/3, \dots, 5\pi/3\}\), \(r_{\text{point}} \sim \sqrt{U(0, 1)}\). Then, the first two coordinates are: \(X_1 = r(z)r_{\text{point}}\cos(\theta)\), and \(X_2 = r(z)r_{\text{point}}\sin(\theta)\).

For rectangular and triangular pyramids, the remaining dimensions \(X_4\) to \(X_{p-1}\), and for star-based pyramids \(X_3\) to \(X_{p-1}\), are treated as noise.

Finally, for the Sierpinski-like pyramid, \texttt{gen\_pyrfrac(n,\ p)} (Figure \ref{fig:pyr-proj} d), let \(X_1, X_2, \dots, X_p\) denote the coordinates of the generated points. The generation process begins with an initial point \(T_0 \in [0, 1]^p\) drawn from a uniform distribution: \(T_0 \sim U(0, 1)^p\). Let \(C_1, C_2, \dots, C_{p+1}\) denote the corner vertices of a \(p\text{-}D\) simplex. At each iteration \(i = 1, \dots, n\), a new point is computed by taking the midpoint between the previous point \(T_{i-1}\) and a randomly selected vertex \(C_k\): \(T_i = 1/2(T_{i-1} + C_k), \quad C_k \in \{C_1, \dots, C_{p+1}\}\). This recursive midpoint rule generates self-similar patterns with systematic voids (holes) between clusters of points. The points remain bounded inside the convex hull of the simplex. The final output is a \(n \times p\) matrix where each row represents a point: \(X = \{T_1, T_2, \dots, T_n\}, \quad X \in \mathbb{R}^{n \times p}\).

Pyramid structures mimic tapering or layered geometries seen in architecture, crystals, and fractal-like natural patterns \citep{kirkby1983}.

\subsubsection{S-curve}\label{s-curve}

The S-curve is a smooth, non-linear manifold in \(3\text{-}D\) space. Using \texttt{gen\_scurve(n)}, it is defined by \(X_1 = \sin(\theta), \quad X_2 \sim U(0, 2), \quad X_3 = \text{sign}(\theta)(\cos(\theta) - 1), \quad \theta \sim U(-3\pi/2, 3\pi/2).\)

This follows the \texttt{s\_curve()} function from snedata \citep{james2025}, itself adapted from \emph{scikit-learn}, but differs by returning a tibble with standardized names (\texttt{x1}, \texttt{x2}, \texttt{x3}), excluding the color variable, and omitting built-in noise (which can be added separately). S-curve is commonly used in manifold learning and dimension reduction as benchmarks for unfolding curved structure.

\subsubsection{Sphere}\label{sphere}

Sphere-shaped structures are useful for evaluating how dimension reduction and clustering algorithms handle curved, symmetric manifolds in high-dimensional spaces. Throughout this section, we follow the standard mathematical terminology: a \emph{sphere} refers to the hollow \((p-1)\)-dimensional surface in \(\mathbb{R}^p\), while a \emph{ball} refers to the filled interior region. The functions generate a variety of spherical forms, including simple circles, uniform and hollow spheres, grid-based spheres, and complex arrangements like clustered spheres within a larger sphere. The first few coordinates define the main geometric form (circle, cycle, sphere, or hemisphere), while higher-dimensional embeddings are achieved by adding noise dimensions. Such spherical or hemispherical structures frequently appear in physical and biological systems, for example in models of celestial bodies, molecular shells, or cell membranes \citep{tinkham2003, alberts2014}.

\begin{figure}[!ht]

{\centering \includegraphics[width=0.8\linewidth,alt={A 3D scatterplot shows a projection of points sampled from a 4‑dimensional hemisphere onto three coordinate axes. Each point represents a location on the surface of a 4‑D unit sphere restricted to one hemisphere. The horizontal (x) and vertical (y) axes, along with depth (z), span roughly from –1 to 1, reflecting the Cartesian coordinates of the projected points. The dots are uniformly scattered over a curved, bowl-like region rather than filling a full sphere, with no strong clusters or gaps, illustrating that the sampling covers only one side of the sphere. If color is used, it likely encodes an additional coordinate or density but does not introduce distinct categorical groups.}]{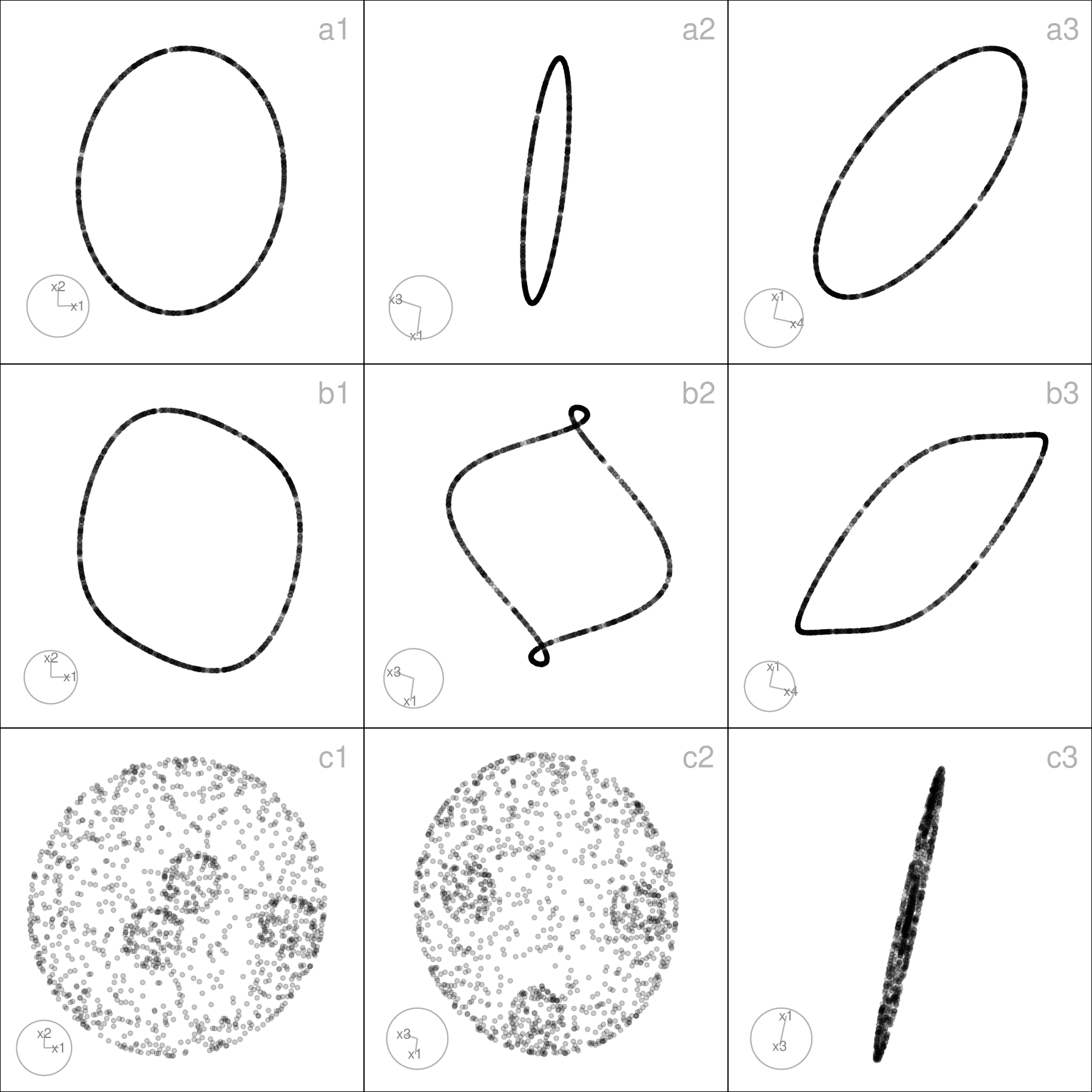} 

}

\caption{Three $2\text{-}D$ projections from $4\text{-}D$, \texttt{circle} (a1-a3), \texttt{curvycycle} (b1-b3), and, $3\text{-}D$ \texttt{clusteredspheres} (c1-c3). The \texttt{circle} structure forms a smooth, closed loop, while \texttt{curvycycle} shows a wavy, continuous pattern forming a twisted ring. The \texttt{clusteredspheres} dataset displays multiple compact spherical groups that are clearly separated in higher dimensions but overlap slightly in some $2\text{-}D$ projections, highlighting how projection can distort spatial relationships. These projections show how simple cyclic, wavy curvilinear, and clustered structures appear in $2\text{-}D$, emphasizing the effects of projection on density, continuity, and separation}\label{fig:sphere-proj}
\end{figure}

The simplest case, \texttt{gen\_circle(n,\ p)} creates a unit circle in two dimensions, with the remaining dimensions forming sinusoidal extensions of the angular parameter at progressively smaller scales (Figure \ref{fig:sphere-proj} a). Let a latent angle variable \(\theta \sim U(0, 2\pi)\). Coordinates in the first two dimensions represent a perfect circle on the plane: \[X_1 = \cos(\theta), \quad X_2 = \sin(\theta).\] For dimensions \(X_3\) through \(X_p\), sinusoidal transformations of the angle \(\theta\) are introduced. The first component is a scaling factor that decreases with the dimension index, defined as \(\text{s}_j = \sqrt{(0.5)^{j-2}}\) for \(j = 3, \dots, p\). The second component is a phase shift that is proportional to the dimension index, specifically designed to decorrelate the curves, given by the formula \(\phi_j = (j - 2)\pi/2p\). Each additional dimension is computed as: \(X_j = \text{s}_{j}\sin(\theta + \phi_j), \quad j = 3, \dots, p\).

For the one-dimensional nonlinear cycle embedded in \(p\text{-}D\) space, \texttt{gen\_curvycycle(n,\ p)} (Figure \ref{fig:sphere-proj} b), let a latent angle variable \(\theta \sim U(0, 2\pi)\). The first three dimensions define a non-circular closed curve, referred to as a \texttt{"curvy\ cycle"}. In this configuration, \(X_1 = \cos(\theta)\) represents horizontal oscillation, while \(X_2 = \sqrt{3}/3 + \sin(\theta)\) introduces a vertical offset to avoid centering the curve at the origin. Additionally, \(X_3 = 1/3\cos(3\theta)\) introduces a third harmonic perturbation that intricately folds the curve three times along its path, creating a unique and complex shape that oscillates in both dimensions while incorporating the effects of the harmonic perturbation.

Together, these define a periodic, non-trivial, closed curve in \(3\text{-}D\) with internal folds that produce a more complex geometry than a standard circle or ellipse. For dimensions \(X_4\) through \(X_p\), additional structured variability is introduced through decreasing amplitude scaling and phase-shifted sine waves. The scaling factor is defined as \(\text{s}_j = \sqrt{(0.5)^{j-3}}\) for \(j\) ranging from \(4\) to \(p\), which means that the amplitude decreases as the dimension increases. Each dimension \(X_j\) is then calculated using the formula \(X_j = \text{s}_j\sin(\theta + \phi_j)\), where the phase shift \(\phi_j\) is given by \(\phi_j = (j - 2)\pi/(2p)\).

Building on simple circular structures, the \texttt{gen\_unifsphere(n,\ r)} function extends the idea to three dimensions by generating \(n\) observations approximately uniformly distributed on the surface of a sphere of radius \(r\). Each observation is computed from spherical coordinates, with \(u \sim U(-1, 1)\) representing \(\cos(\phi)\) and \(\theta \sim U(0, 2\pi)\) the azimuthal angle. Cartesian coordinates are then defined as \[X_1 = r\sqrt{1 - u^2}\cos(\theta), \quad X_2 = r\sqrt{1 - u^2}\sin(\theta),\text{ and }X_3 = ru,\] ensuring uniform distribution on the surface (not within) of the sphere.

In contrast, the \texttt{gen\_hollowsphere(n,\ p)} function, a wrapper around \texttt{geozoo::sphere.hollow()}, generates \(n\) points uniformly distributed only on the surface of the \((p-1)\)-dimensional sphere embedded in \(\mathbb{R}^p\). This results in a hollow shell-like structure with no interior points. For example, when \(p=3\), \texttt{gen\_unifsphere()} produces a solid ball in \(3\text{-}D\) space, whereas \texttt{gen\_hollowsphere()} produces only the spherical boundary. These paired structures allow controlled experiments to investigate how algorithms behave when data is concentrated throughout the full volume versus constrained to the boundary.

In addition, the \texttt{gen\_gridedsphere(n)} function constructs a \(p\)-dimensional dataset consisting of approximately \(n\) points arranged on the surface of the unit \((p-1)\)-sphere embedded in \(\mathbb{R}^p\) (Figure \ref{fig:sphere-proj} d). Rather than sampling points uniformly, this function creates a deterministic grid in spherical coordinates, using \((p-1)\) angular variables: the first \((p-2)\) angles are taken from \([0, \pi]\), and the final angle from \([0, 2\pi]\). The number of grid points along each angular dimension is determined by decomposing \(n\) into \((p-1)\) approximately equal integer factors via \texttt{gen\_nproduct(n,\ p\ -\ 1)}.

Each grid point is subsequently mapped into Cartesian space via the standard hyperspherical-to-Cartesian transformation,

\[
\begin{aligned}
X_1 &= \cos(\theta_1), \\
X_2 &= \sin(\theta_1)\cos(\theta_2), \\
X_3 &= \sin(\theta_1)\sin(\theta_2)\cos(\theta_3), \\
&\;\;\vdots \\
X_{p-1} &= \sin(\theta_1)\sin(\theta_2)\cdots \sin(\theta_{p-2})\cos(\theta_{p-1}), \\
X_p &= \sin(\theta_1)\sin(\theta_2)\cdots \sin(\theta_{p-2})\sin(\theta_{p-1}).
\end{aligned}
\]

The result is a deterministic grid of points lying exactly on the surface of the unit \((p-1)\)-sphere, without any additional noise dimensions.

For more heterogeneous structures, the \texttt{gen\_clusteredspheres(n,\ k,\ r,\ loc)} function generates one large sphere of radius \(r_1\) and \(k\) smaller spheres of radius \(r_2\), each centered at a different random location (Figure \ref{fig:sphere-proj} e). A large Uniform ball centered at the origin is created by sampling \(n_1\) points uniformly on the surface of a \(p\text{-}D\) sphere with a radius of \(r_1\). The sampling is executed using the function \texttt{gen\_unifsphere(n\_1,\ r\_1)}, which generates the desired points in the specified dimensional space. In generation of \(k\) smaller Uniform balls, each sphere contains \(n_2\) points that are sampled uniformly on a sphere with a radius of \(r_2\). These spheres are positioned at distinct random locations in \(p\text{-}D\), with the center of each sphere being drawn from a normal distribution \(N(0, \texttt{loc}^2 I_p)\). Points on spheres are generated using the standard hyperspherical method, which involves sampling \(u \sim U(-1, 1)\) to determine the cosine of the polar angle, and sampling \(\theta \sim U(0, 2\pi)\) to determine the azimuthal angle (for \(3\text{-}D\)). Each observation is classified by cluster, with labels such as ``big'' for the large central sphere and ``small\_1'' to ``small\_k'' for the smaller spheres.

Finally, the \texttt{gen\_hemisphere(n,\ p)} function restricts sampling to a hemisphere of a \(4\text{-}D\) sphere (Figure \ref{fig:sphere-proj} f). Using spherical coordinates, the azimuthal angle \(\theta_1 \sim U(0, \pi)\) in the \((x_1, x_2)\) plane, while the elevation angle \(\theta_2 \sim U(0, \pi)\) in the \((x_2, x_3)\) plane. Additionally, \(\theta_3 \sim U(0, \pi/2)\) in the \((x_3, x_4)\) plane, ensuring that the points remain restricted to a hemisphere. The coordinates are transformed into \(4\text{-}D\) Cartesian space: \[X_1 = \sin(\theta_1)\cos(\theta_2), \quad X_2 = \sin(\theta_1)\sin(\theta_2), \\\quad X_3 = \cos(\theta_1)\cos(\theta_3), \quad X_4 = \cos(\theta_1)\sin(\theta_3).\] This produces points on one side of a \(4\text{-}D\) unit sphere, effectively generating a \(4\text{-}D\) hemisphere.

\subsubsection{Swiss Roll}\label{swiss-roll}

The Swiss roll is a plane curled into \(3\text{-}D\), and is a commonly used example of a nonlinear manifold. The \texttt{gen\_swissroll(n,\ w)} generates points as \(X_1 = t \cos(t), \quad X_2 = t \sin(t), \quad X_3 \sim U(w_1, w_2), \quad t \sim U(0, 3\pi).\)

Compared with \texttt{snedata::swiss\_roll()} \citep{james2025}, this implementation (i) samples \(t\) over \([0, 3\pi]\) instead of \([1.5\pi, 4.5\pi]\), (ii) allows a flexible vertical range \(w = (w_1, w_2)\) rather than fixing \(z \in [0, z_{\max}]\), and (iii) returns a tibble with \texttt{x1,\ x2,\ x3} instead of adding a color variable.

The Swiss roll is a classic benchmark for manifold learning, illustrating how a curved surface can be ``unrolled'' into lower dimensions. Similar spiral-like forms appear in galaxies, protein folding, and coiled materials \citep{dimitris2002}.

\subsubsection{Trefoil knots}\label{trefoil-knots}

\begin{figure}[!ht]

{\centering \includegraphics[width=0.8\linewidth,alt={The figure shows multiple scatterplots displaying two-dimensional projections of trefoil-shaped data derived from four- and three-dimensional spaces. In panels a1–a3, the trefoil4d data are shown as three different 2-D projections, where the horizontal and vertical axes represent linear combinations of the four original variables. The projected points form smooth, continuous looping curves that intersect and overlap in places, creating knot-like shapes whose apparent crossings and separations vary across projections. In panels b1–b3, the trefoil3d data are shown in three 2-D projections. The points trace a compact, closed loop with a characteristic trefoil-like knot structure, appearing smoother and less spread out than in the four-dimensional case. The figure highlights how the same underlying knot structure appears differently under changes in projection and dimensionality.}]{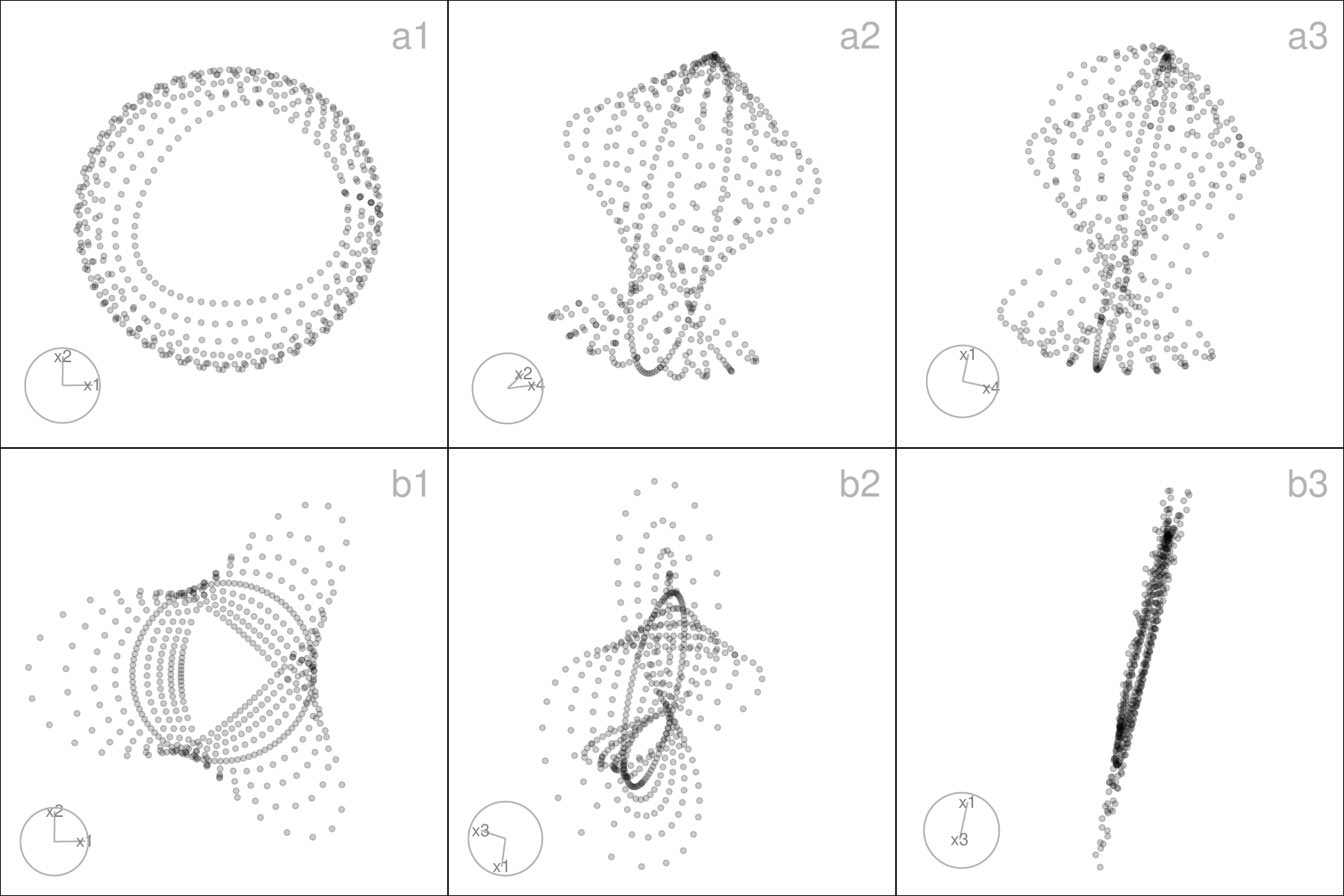} 

}

\caption{Three $2\text{-}D$ projections from $4\text{-}D$, \texttt{trefoil4d} (a1-a3) and $3\text{-}D$ \texttt{trefoil3d} (b1-b3) data. The \texttt{trefoil4d} structure represents a higher-dimensional extension of the classic trefoil knot, revealing complex twisting and looping patterns that remain continuous across projections. In contrast, the \texttt{trefoil3d} dataset maintains a simpler, more compact knot-like form, showing how dimensional extension adds curvature and separation in the embedded space. These projections illustrate a range of looping structures in high-dimensions.}\label{fig:trefoil-proj}
\end{figure}

The Trefoil is a closed, nontrivial one-dimensional manifold embedded in \(3\text{-}D\) or \(4\text{-}D\) space (Figure \ref{fig:trefoil-proj}). The trefoil features topological complexity in the form of self-overlaps, making it a valuable test case for evaluating the ability of non-linear dimension reduction methods to preserve global structure, loops, and embeddings in high-dimensional data.

For the \(4\text{-}D\) trefoil knot \citep{laurent2024}, the function \texttt{gen\_trefoil4d(n,\ steps)} generates the structure on the \(3\)-sphere (\(S^3 \subset \mathbb{R}^4\)) using two angular parameters, \(\theta\) and \(\phi\). A band of thickness around the knot path is controlled by the \texttt{steps} argument, while the number of \(\theta\) and \(\phi\) values is determined by the \texttt{steps} and \texttt{n} arguments, respectively (Figure \ref{fig:trefoil-proj} a). The coordinates are defined as \[X_1 = \cos(\theta) \cos(\phi), \quad X_2 = \cos(\theta) \sin(\phi), \\\quad X_3 = \sin(\theta) \cos(1.5 \phi),\text{ and }X_4 = \sin(\theta) \sin(1.5 \phi),\] where \(\theta\) parameterizes the band thickness and \(\phi\) parameterizes the knot trajectory.

For the \(3\text{-}D\) stereographic projection \citep{laurent2024}, \texttt{gen\_trefoil3d(n,\ steps)} maps each point \((X_1, X_2, X_3, X_4) \in \mathbb{R}^4\) to \((X_1', X_2', X_3') \in \mathbb{R}^3\text{ using }X_1' = X_1 / (1 - X_4), \quad X_2' = X_2 / (1 - X_4),\text{ and }X_3' = X_3 / (1 - X_4),\) excluding points where \(X_4 = 1\) to avoid division by zero (Figure \ref{fig:trefoil-proj} b).

The trefoil knot appears in molecular biology (DNA/protein knotting), fluid dynamics (knotted vortices), and physics (topological phases), making it a useful benchmark for testing whether dimension reduction preserves global loops and topology \citep{witten1985, arsuaga2002}.

\subsubsection{Trigonometric}\label{trigonometric}

\begin{figure}[!ht]

{\centering \includegraphics[width=0.7\linewidth,alt={The figure consists of four rows of panels, each displaying three separate 2-D projection scatterplots from a four-dimensional synthetic dataset. Panels (a1–a3) show the curvycylinder dataset, where points form elongated, band-like structures that bend smoothly across projections, appearing cylindrical with gentle nonlinear curvature. Panels (b1–b3) show the sphericalspiral dataset, with points tracing curved, looping paths that resemble spiral arcs on a rounded surface, changing orientation across projections. Panels (c1–c3) show the conicspiral dataset, where point clouds expand outward in some projections while twisting along curved trajectories, creating tapered, spiral-like shapes. Panels (d1–d3) show the nonlinear dataset, in which points form warped surfaces with alternating tight and spread regions, visible as oscillating bands and smooth waves depending on the projection. Across all datasets, point density, curvature, and apparent continuity vary between projections, illustrating how different geometric structures appear under multiple 2-D views of the same four-dimensional data.}]{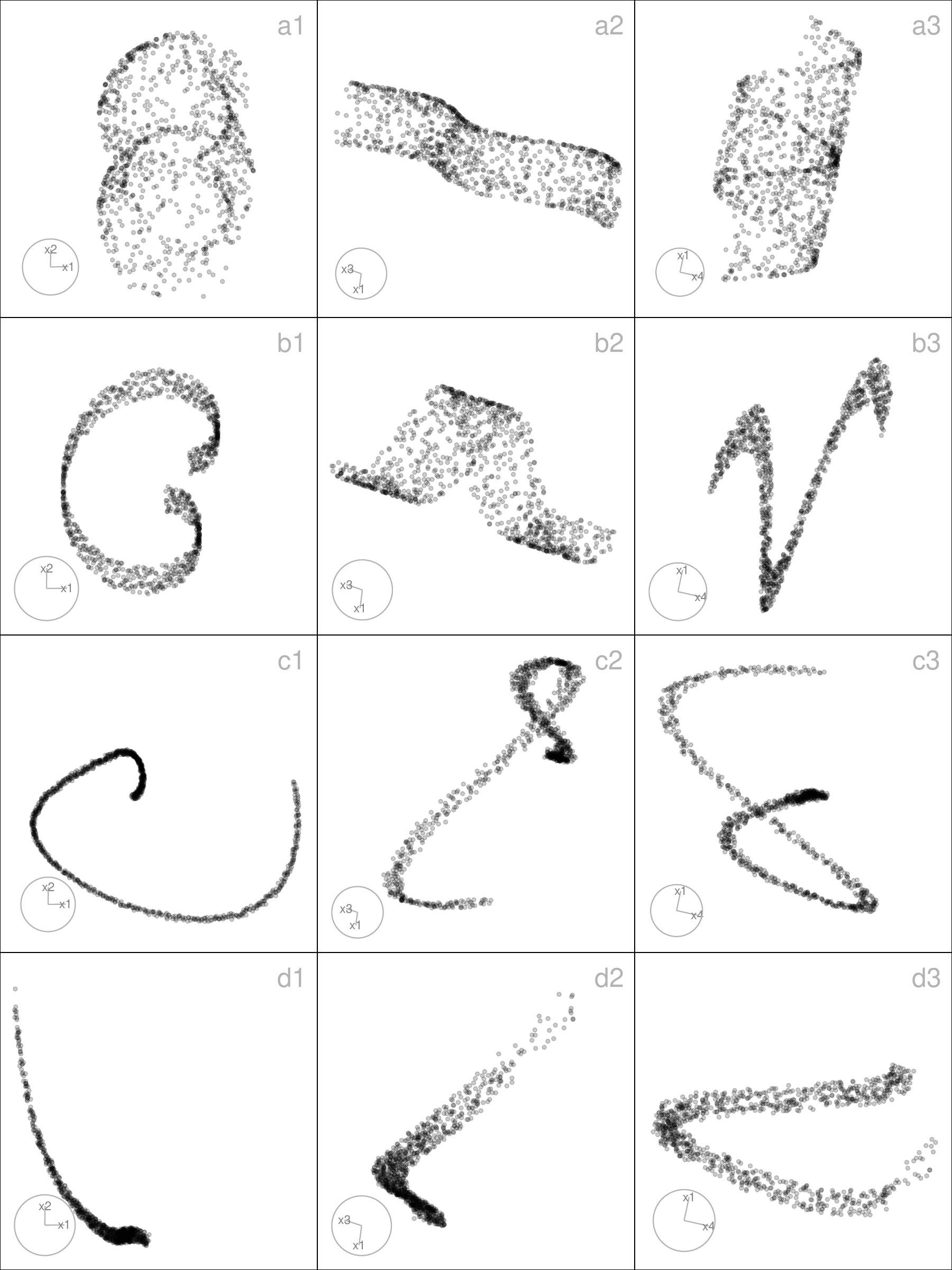} 

}

\caption{Three $2\text{-}D$ projections from $4\text{-}D$, for the \texttt{curvycylinder} (a1-a3), \texttt{sphericalspiral} (b1-b3), \texttt{conicspiral} (c1-c3), and \texttt{nonlinear} (d1-d3) data. The \texttt{curvycylinder} shows a cylindrical manifold with a nonlinear twist along its height, producing smooth, continuous curvature. The \texttt{sphericalspiral} forms a spiral path on a spherical surface, combining circular and vertical motion in a helical form. The \texttt{conicspiral} spreads radially while ascending, forming a conical helix with twisting variations in a non-radial dimension. The \texttt{nonlinear} dataset exhibits a warped $2\text{-}D$ surface with sharp oscillations and smooth waves, reflecting complex nonlinear interactions. Each shows variations in curvature, density, and continuity.}\label{fig:triginometric-proj}
\end{figure}

Trigonometric-based structures provide flexible ways to simulate complex curved patterns and spirals that often arise in real-world high-dimensional data, such as in biological trajectories, or physical systems (Figure \ref{fig:triginometric-proj}). The main geometry is defined by the first few coordinates: crescents (\(p=2\)), cylinders, spirals, and helices (\(p=4\)). These structures are particularly valuable for testing how well dimension reduction and clustering algorithms preserve intricate geometric and topological features \citep{calladine1997, gershenfeld2000}.

First, the \texttt{gen\_crescent(n,\ p)} function generates a \(p\)-dimensional dataset of \(n\) observations based on a \(2\text{-}D\) crescent-shaped manifold with optional structured high-dimensional noise (Figure \ref{fig:triginometric-proj} a). Let \(\{\theta_i\}_{i=1}^n\) be a sequence of \(n\) evenly spaced angles on the interval \([\pi/6, 2\pi]\), defined as \(\theta_i = \frac{\pi}{6} + (i-1)\frac{2\pi - \pi/6}{n-1}, \quad i = 1,\dots,n\). The corresponding \(2\text{-}D\) coordinates are defined by: \[X_{i1} = \cos(\theta_i), \quad X_{i2} = \sin(\theta_i).\]

Second, the \texttt{gen\_curvycylinder(n,\ p,\ h)} function generates a \(p\text{-}D\) dataset of \(n\) observations structured as a \(3\text{-}D\) cylindrical manifold with an added nonlinear curvy dimension, and optional noise dimensions when \(p > 4\) (Figure \ref{fig:triginometric-proj} b). The core structure consists of a circular base and height values, extended by a nonlinear fourth dimension. Let \(\theta \sim U(0, 3\pi)\) represent a random angle on a circular base and \(z \sim U(0, h)\) represent the height along the cylinder. The coordinates are defined as: \(X_1 = \cos(\theta)\) (Circular base, x-axis), \(X_2 = \sin(\theta)\) (Circular base, y-axis), \(X_3 = z\) (Linear height), and \(X_4 = \sin(z)\) (Nonlinear curvy variation along height).

For a spiraling path on a spherical surface in the first four dimensions, \texttt{gen\_sphericalspiral(n,\ p,\ spins)} (Figure \ref{fig:triginometric-proj} c), let \(\theta \in [0, 2\pi \times \text{spins}]\) be the azimuthal angle (longitude), controls the number of spiral turns and the \(\phi \in [0, \pi]\) be the polar angle (latitude), controls the vertical sweep from the north to the south pole. Cartesian coordinates from spherical conversion: \(X_1 = \sin(\phi)\cos(\theta)\), \(X_2 = \sin(\phi)\sin(\theta)\), \(X_3 = \cos(\phi) + \varepsilon\), where \(\varepsilon \sim U(-0.5, 0.5)\) introduces vertical jitter, and \(X_4 = \theta / \max(\theta)\): a normalized progression along the spiral path. This generates a spherical spiral curve embedded in \(4\text{-}D\) space, combining both circular and vertical movement, with gentle curvature and non-linear progression.

For a helical spiral in four dimensions, \texttt{gen\_helicalspiral(n,\ p)} (Figure \ref{fig:triginometric-proj} d), let \(\theta \in [0, 5\pi/4]\) be a sequence of angles controlling rotation around a circle. Cartesian coordinates; \(X_1 = \cos(\theta)\): circular trajectory along the x-axis, \(X_2 = \sin(\theta)\): circular trajectory along the y-axis, \(X_3 = 0.05\theta + \varepsilon_3\), with \(\varepsilon_3 \sim U(-0.5, 0.5)\): linear progression (height) with vertical jitter, simulating a helix, and \(X_4 = 0.1\sin(\theta)\): oscillates with \(\theta\), representing a periodic ``wobble'' along the fourth dimension.

Similarly, the \texttt{gen\_conicspiral(n,\ p,\ spins)} function generates a dataset of \(n\) points forming a conical spiral in the first four dimensions of \(p\text{-}D\) (Figure \ref{fig:triginometric-proj} e). The geometry combines radial expansion, vertical elevation, and spiral deformation, simulating a structure that fans out like a \(3\text{-}D\) conic helix. The shape is defined by parameter \(\theta \in [0, 2\pi\text{spins}]\), controlling the angular progression of the spiral. The Archimedean spiral in the horizontal plane is represented by; \(X_1 = \theta\cos(\theta)\) for radial expansion in \(x\), and \(X_2 = \theta\sin(\theta)\) for radial expansion in \(y\). The growth pattern resembles a cone, with the height increasing according to \(X_3 = 2\theta / \max(\theta) + \varepsilon_3\), with \(\varepsilon_3 \sim U(-0.1, 0.6).\) Spiral modulation in the fourth dimension is represented by \(X_4 = \theta\sin(2\theta) + \varepsilon_4\), with \(\varepsilon_4 \sim U(-0.1, 0.6)\) which simulates a twisting helical component in a non-radial dimension.

Finally, the \texttt{gen\_nonlinear(n,\ p,\ hc,\ non\_fac)} function simulates a nonlinear \(2\text{-}D\) surface embedded in higher dimensions, constructed using inverse and trigonometric transformations applied to independent variables (Figure \ref{fig:triginometric-proj} f). The \(X_{1} \sim U(0.1, 2)\): base variable (avoids zero to prevent division errors), \(X_{3} \sim U(0.1, 0.8)\): independent auxiliary variable, \(X_{2} = hc/X_{1} + \text{non\_fac}\sin(X_{1})\): nonlinear combination of hyperbolic and sinusoidal transformations, creating sharp curvature and oscillation, and \(X_{4} = \cos(\pi X_{1}) + \varepsilon\), with \(\varepsilon \sim U(-0.1, 0.1)\): additional nonlinear variation based on cosine, simulating more subtle periodic structure. These transformations together result in a nonlinear surface warped in multiple ways: sharp vertical shifts due to inverse terms, smooth waves from sine and cosine, and additional jitter.

\subsection{Generate a spherical or hyperspherical hole within a structure}\label{generate-a-spherical-or-hyperspherical-hole-within-a-structure}

The package provides functionality for generating datasets with spherical hole (in \(2\text{-}D\)/\(3\text{-}D\)) or, more generally, hyperspherical hole (in higher dimensions). These structures are valuable for evaluating how dimension reduction methods and clustering algorithms handle incomplete manifolds or missing regions of the data space. A hyperspherical hole introduces topological complexity: the structure remains continuous but contains excluded regions (voids) that algorithms must correctly represent in lower-dimensional embeddings.

The core function \texttt{gen\_hole(df,\ anchor,\ r)} removes points from a dataset that fall within a user-specified hypersphere. Formally, given data points (\(x \in \mathbb{R}^p\)), a center (\(a \in \mathbb{R}^p\)), and radius (\(r > 0\)), only points satisfying \(\sqrt{\sum_{i=1}^n(x_i-a_i)^2} > r\) are retained. The anchor point (\(a\)) can either be user-specified or default to the dataset mean, and radius (\(r\)) is controlled by the user, with safeguards to avoid trivial or degenerate cases. Because it operates generically on any dataset, spherical or hyperspherical holes can be embedded in a wide range of geometric structures.

Two specialized wrappers illustrate this idea. The function \texttt{gen\_scurvehole(n,\ r\_hole)} generates an S-curve with a spherical hole by applying \texttt{gen\_hole()} to the output of \texttt{gen\_scurve()}. This structure has been used in prior diagnostic studies of NLDR methods \citep[\citet{van2007}]{tenenbaum2000}, since it tests the ability of algorithms to capture non-linear manifolds that are not simply connected. The second wrapper, \texttt{gen\_unifcubehole(n,\ p,\ r\_hole)}, generates uniformly sampled cube data with a hyperspherical hole. By embedding a hyperspherical void inside a convex high-dimensional structure, this creates non-convex regions that challenge algorithms in terms of separability and neighborhood preservation.

\subsection{Generate noise dimensions}\label{generate-noise-dimensions}

High-dimensional data structures often benefit from the addition of auxiliary noise dimensions, which can be used to assess the robustness of dimension reduction and clustering algorithms. The functions in this section provide flexible ways to generate random noise dimensions, ranging from purely random Gaussian variables to more structured, wavy patterns that mimic nonlinear distortions in high-dimensional space. These functions can be applied independently or combined with other geometric structures to create complex simulated datasets. Table \ref{tab:noise-tb-pdf} details these functions.

\begin{table}

\caption{\label{tab:noise-tb-pdf}cardinalR noise dimensions generation functions}
\centering
\begin{tabular}[t]{>{\raggedright\arraybackslash}p{2.5cm}>{\raggedright\arraybackslash}p{10.5cm}}
\toprule
Function & Explanation\\
\midrule
\texttt{gen\_noisedims} & Gaussian noise dimensions with optional mean and standard deviation.\\
\texttt{gen\_wavydims1} & Wavy noise dimensions based on a user-specified theta sequence with added jitter.\\
\texttt{gen\_wavydims2} & Wavy noise dimensions using polynomial transformations of an existing dimension vector.\\
\texttt{gen\_wavydims3} & Wavy noise dimensions using a combination of polynomial and sine transformations based on the first three dimensions of a dataset.\\
\bottomrule
\end{tabular}
\end{table}

The \texttt{gen\_noisedims(n,\ p,\ m,\ s)} function generates \(p\) independent Gaussian noise dimensions,

\[
X_j \sim N(m_j, s_j^2), \quad j = 1, \dots, p,
\]

with odd-numbered dimensions multiplied by \(-1\). This does not affect independence, since all noise dimensions are generated independently. The sign alternation is included only to avoid consistent directional drift and to ensure a symmetric appearance of noise when visualized or combined with other simulated structures.

For scenarios where noise should follow a smooth wavy pattern, \texttt{gen\_wavydims1(n,\ p,\ theta)} generates dimensions as

\[
X_j = \alpha_j \theta + \varepsilon_j, \quad \varepsilon_j \sim N(0, \sigma^2), \quad j = 1, \dots, p,
\]

where each dimension is scaled by a different factor \(\alpha_j\), producing structured noise that oscillates along the latent parameter \(\theta\), mimicking trends or trajectories observed in real-world data.

The \texttt{gen\_wavydims2(n,\ p,\ x\_1)} function extends this approach by applying a nonlinear transformation to an existing dimension vector \(x_1\):

\[
X_j = \beta_j \, (-1)^{\lfloor j/2 \rfloor} \, x_1^{k_j} + \varepsilon_j, \quad j = 1, \dots, p,
\]

where \(k_j\) is a randomly chosen polynomial power, \(\beta_j\) is a scaling factor, and \(\varepsilon_j\) is small uniform noise.

Finally, \texttt{gen\_wavydims3(n,\ p,\ data)} generates noise for datasets with multiple correlated dimensions. The first three dimensions are small perturbations of the original coordinates \((X_1, X_2, X_3)\), while higher dimensions are constructed via nonlinear combinations, including polynomial and trigonometric transformations, e.g.,

\[
X_j = f_j(X_1, X_2, X_3) + \varepsilon_j, \quad j > 3,
\]

producing high-dimensional noise that preserves some geometric correlation with the base structure while introducing additional complexity.

\subsection{Rotating shape generators}\label{rotating-shape-generators}

In \(p\text{-}D\) space, a rotation is an orthogonal transformation that changes the orientation of data while preserving its total variance and pairwise distances. The function \texttt{gen\_rotation()} generates such rotation matrices for any dimension, given a list of rotation planes (axis pairs) and angles.

\subsection{Multiple cluster examples}\label{multiple-cluster-examples}

By using the shape generators mentioned above, we can create various examples of multiple clusters. The package includes some of these examples, which are described in Table \ref{tab:odd-shape-tb-pdf}.

\begin{table}

\caption{\label{tab:odd-shape-tb-pdf}cardinalR multiple clusters generation functions}
\centering
\begin{tabular}[t]{>{\raggedright\arraybackslash}p{3.5cm}>{\raggedright\arraybackslash}p{8.5cm}}
\toprule
Function & Explanation\\
\midrule
\texttt{make\_mobiusgau} & Möbius-like cluster combined with a Gaussian.\\
\texttt{make\_multigau} & Multiple Gaussian clusters in high-dimensional space.\\
\texttt{make\_curvygau} & Curvilinear cluster with a Gaussian cluster.\\
\texttt{make\_klink\_circles} & K-link circular clusters (non-linear circular patterns).\\
\texttt{make\_chain\_circles} & Chain-like circular clusters connected sequentially.\\
\texttt{make\_klink\_curvycycle} & K-link curvy cycle clusters (curvilinear loop structures).\\
\texttt{make\_chain\_curvycycle} & Chain-like curvy cycle clusters connected sequentially.\\
\texttt{make\_gaucircles} & Circular clusters with a Gaussian cluster in the middle.\\
\texttt{make\_gaucurvycycle} & Curvy circular clusters with a Gaussian in the middle.\\
\texttt{make\_onegrid} & Single grid in two dimensions.\\
\texttt{make\_twogrid\_overlap} & Two overlapping grids.\\
\texttt{make\_twogrid\_shift} & Two grids shifted relative to each other.\\
\texttt{make\_shape\_para} & Parallel shaped clusters.\\
\bottomrule
\end{tabular}
\end{table}

\subsection{Additional functions}\label{additional-functions}

The package includes various supplementary tools in addition to the shape generating functions mentioned earlier. These tools allow users to create background noise, randomize the rows of the data, relocate clusters, generate a vector whose product and sum are approximately equal to a target value, rotate structures, and normalize the data. Table \ref{tab:add-tb-pdf} details these functions. More detailed explanations are available in \href{https://jayanilakshika.github.io/cardinalR/articles/03additionalfun.html}{jayanilakshika.github.io/cardinalR/articles/03additionalfun}.

\begin{table}

\caption{\label{tab:add-tb-pdf}cardinalR additional functions}
\centering
\begin{tabular}[t]{>{\raggedright\arraybackslash}p{4cm}>{\raggedright\arraybackslash}p{8cm}}
\toprule
Function & Explanation\\
\midrule
\texttt{gen\_bkgnoise} & Adds background noise.\\
\texttt{randomize\_rows} & Randomizes the rows of input data.\\
\texttt{relocate\_clusters} & Relocates the clusters.\\
\texttt{gen\_nproduct} & Generates a vector of positive integers whose product is approximately equal to a target value.\\
\texttt{gen\_nsum} & Generates a vector of positive integers whose summation is approximately equal to a target value.\\
\texttt{normalize\_data} & Normalizes data.\\
\bottomrule
\end{tabular}
\end{table}

\section{Application}\label{application}

This section demonstrates how the package can be used to generate complex high-dimensional datasets, evaluate dimension reduction (DR) and clustering methods. The example shows how diverse geometric structures can be simulated and analyzed to assess algorithmic behavior.

\begin{figure}[!ht]
\includegraphics[width=1\linewidth,alt={The figure shows three 2-D projection plots of synthetic 4-D data containing five clusters with distinct geometric structures. The horizontal and vertical axes represent two projected dimensions, and no explicit units are shown. Points are colored by cluster, with distinct colors representing the helical spiral, hemisphere, uniform cube, cone, and Gaussian clusters. The helical spiral cluster appears as an elongated, twisted band; the hemisphere cluster forms a curved, arc-like concentration; the uniform cube cluster fills a roughly rectangular region with fairly even density; the cone cluster exhibits a density gradient, narrowing at one end and spreading out at the other; and the Gaussian cluster forms a compact, approximately circular group near the centre of the plot. The clusters appear very close to one another in these projections, illustrating the difficulty of distinguishing cluster structure when projections bring clusters into close proximity.}]{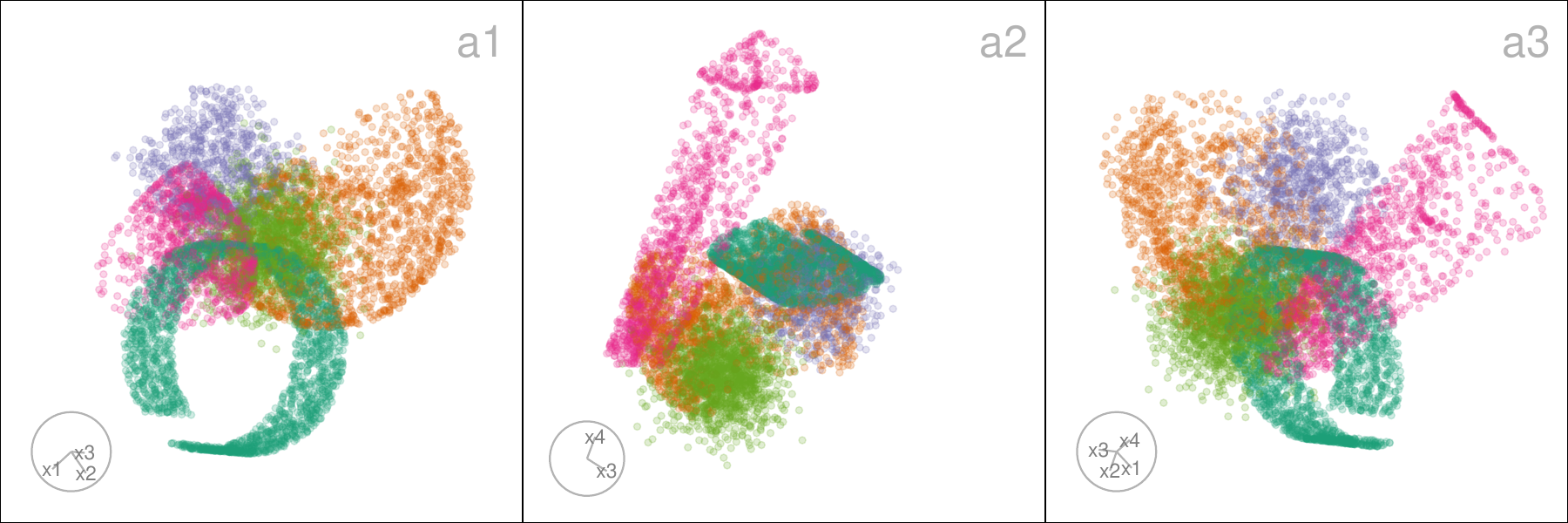} \caption{Three $2\text{-}D$ projections from $4\text{-}D$, for the five clusters data. The helical spiral cluster is represented in dark green, the hemisphere cluster in orange, the uniform cube-shaped cluster in purple, the blunted cone cluster in pink, and the Gaussian-shaped cluster in light green.}\label{fig:highd-proj}
\end{figure}

To illustrate how high-dimensional clustered data can be generated using \texttt{cardinalR}, we generate a dataset with five clusters in \(4\text{-}D\), each representing distinct geometric characteristics: a \emph{helical spiral} (elongated and twisted), a \emph{hemisphere} (curved surface), a \emph{uniform cube} (isotropic distribution), a \emph{cone} (density gradient), and a \emph{Gaussian} cluster (compact and spherical) (Figure \ref{fig:highd-proj}). Each cluster has a unique number of points and scaling factor, representing variation in cluster size and spread across the \(4\text{-}D\) space.

\begin{verbatim}
positions <- geozoo::simplex(p=4)$points
positions <- positions * 0.3

five_clusts <- gen_multicluster(n = c(2250, 1500, 750, 1250, 1750), k = 5,
                       loc = positions,
                       scale = c(0.25, 0.35, 0.3, 1, 0.3),
                       shape = c("helicalspiral", "hemisphere", "unifcube", 
                                 "cone", "gaussian"),
                       rotation = NULL,
                       is_bkg = FALSE)
\end{verbatim}

\subsection{Evaluating dimension reduction (DR) methods}\label{evaluating-dimension-reduction-dr-methods}

We applied six popular DR techniques to the generated dataset: Principal Component Analysis (PCA) \citep{jolliffe2011}, tSNE, uniform manifold approximation and projection (UMAP) \citep{leland2018}, potential of heat-diffusion for affinity-based trajectory embedding (PHATE) algorithm \citep{moon2019}, large-scale dimensionality reduction Using triplets (TriMAP) \citep{amid2019}, and pairwise controlled manifold approximation (PaCMAP) \citep{yingfan2021}.

\begin{figure}[!ht]
\includegraphics[width=1\linewidth,alt={A multi-panel figure compares 2-D scatterplots from six nonlinear dimensionality reduction methods applied to the same dataset with five true clusters. Each panel plots the first embedding dimension on the horizontal axis and the second embedding dimension on the vertical axis. Points represent individual observations and are colored by cluster membership, with six distinct cluster colors reused consistently across panels. In the tSNE panel, clusters form compact, clearly separated groups with small gaps between clusters, indicating strong preservation of both local neighbourhoods and the global cluster layout. UMAP and PaCMAP also show six visibly distinct clusters that are moderately well separated but with slightly more overlap and distortion than tSNE. PHATE produces curved, nonlinear cluster shapes where clusters are stretched or intertwined, obscuring the original simple cluster geometry. TriMAP collapses the data into three main visible groups instead of six, with only small distances between these groups, suggesting loss of finer cluster structure. PCA displays the weakest structure: clusters overlap substantially and align along a roughly linear or planar trend, failing to reflect the underlying non-linear separation among the six groups.}]{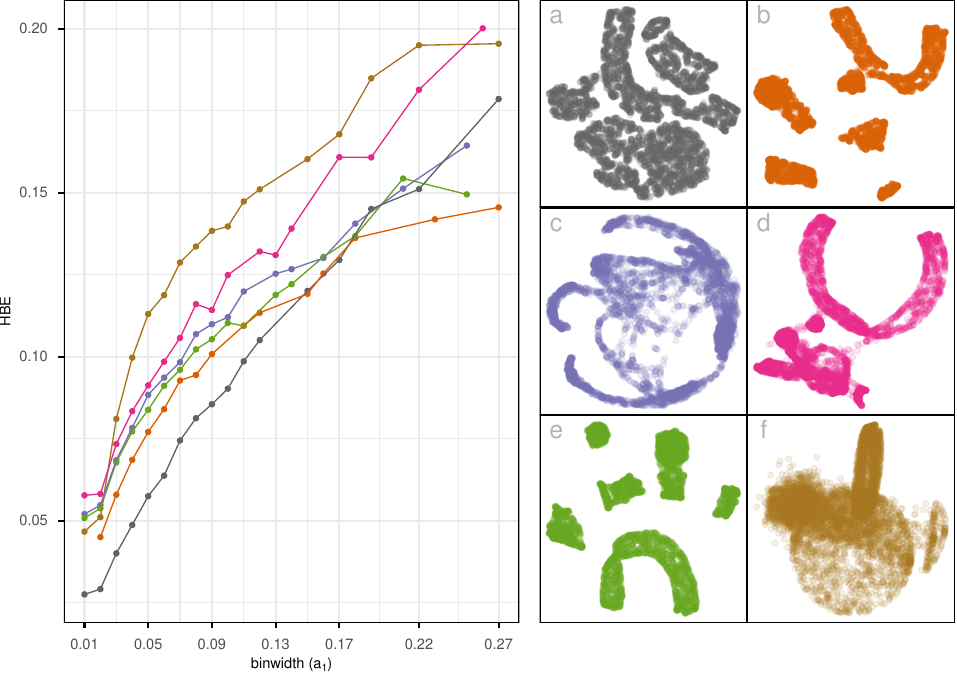} \caption{Assessing which of the 6 NLDR layouts ((a) tSNE, (b) UMAP, (c) PAHTE, (d) TriMAP, (e) PaCMAP, and (f) PCA) of the five clusters data is the better representation using HBE for varying binwidth ($a_1$). Color is used for the lines and points in the left plot to match the scatterplots of the NLDR layouts (a-f). Layout f is universally poor. Layouts a and b are universally optimal. Layout b shows six well-separated clusters and layout a shows close clusters, thus layout a is the best choice.}\label{fig:fig-nldr-layouts}
\end{figure}

To assess their performance, we computed the hexbin error (HBE) between the observed high-dimensional data and the fitted values, defined as the high-dimensional mappings of the bin centroids \citep{gamage2025c}. A lower HBE indicates that the method better preserves the high-dimensional structure in its low-dimensional embedding.

As shown in Figure \ref{fig:fig-nldr-layouts}, tSNE (Figure \ref{fig:fig-nldr-layouts} a) achieved the lowest HBE across bin widths (mostly tiny), indicating high preservation of both local and global structures. Its layout displays well-separated clusters with minimal inter-cluster distances, making it the most faithful representation of the underlying data structure. UMAP and PaCMAP (Figure \ref{fig:fig-nldr-layouts} b and e) produced moderately accurate embeddings, although the six clusters appear more well-separated, while PHATE (Figure \ref{fig:fig-nldr-layouts} c) show nonlinear cluster structures irrespective of the original structure. Also, TriMAP (Figure \ref{fig:fig-nldr-layouts} d) has high HBE, and show three clusters with small distances. PCA (Figure \ref{fig:fig-nldr-layouts} f) failed to capture the non-linear geometry, leading to the highest HBE.

\subsection{Benchmarking clustering algorithms}\label{benchmarking-clustering-algorithms}

To further evaluate the structure of the generated data, we benchmarked three clustering algorithms: \textbf{\(k\)-means} \citep[Chapter 20 of][]{boehmke2019}, \textbf{hierarchical} \citep{murtagh2012}, and \textbf{model-based clustering} \citep{chris2002, scrucca2023} using the simulated dataset. Model-based clustering performed the \texttt{"EII"} covariance structure. Under this parameterization, clusters are spherical with equal volume and equal shape, and no orientation parameter is estimated. Cluster validity statistics were computed using the \texttt{cluster.stats()} function from the \texttt{fpc} package \citep{christian2024}.

\begin{figure}[!]
\includegraphics[width=1\linewidth,alt={Multi-panel line chart showing six cluster quality metrics across numbers of clusters (x-axis: 2 to 10 clusters) for three clustering methods (k-means, hierarchical, model-based). Each panel displays one metric with the y-axis scaled to that metric’s values (not explicitly labeled in the text, but each metric is treated as higher-is-better or lower-is-better as described). Within each panel, three lines (one per method) trace the metric value as the number of clusters increases. For Pearson gamma, values for all three methods rise steeply up to about 5 clusters and then level off. For the Calinski–Harabasz index, values increase sharply from 4 to 5 clusters. For Dunn, the k-means line peaks around 4 clusters, while the hierarchical and model-based lines peak around 5. For WB ratio and within-cluster sum of squares, all three methods show a generally monotonic decline as the number of clusters increases, with a visible bend or elbow around 5 clusters. For the S-index, the k-means line reaches its best value around 4 clusters, the hierarchical line around 3, 6, or 8 clusters, and the model-based line around 4 or 8. Across panels, the k-means line is typically at or near the most favorable values for each metric, and several metrics simultaneously favor solutions with about 4–5 clusters.}]{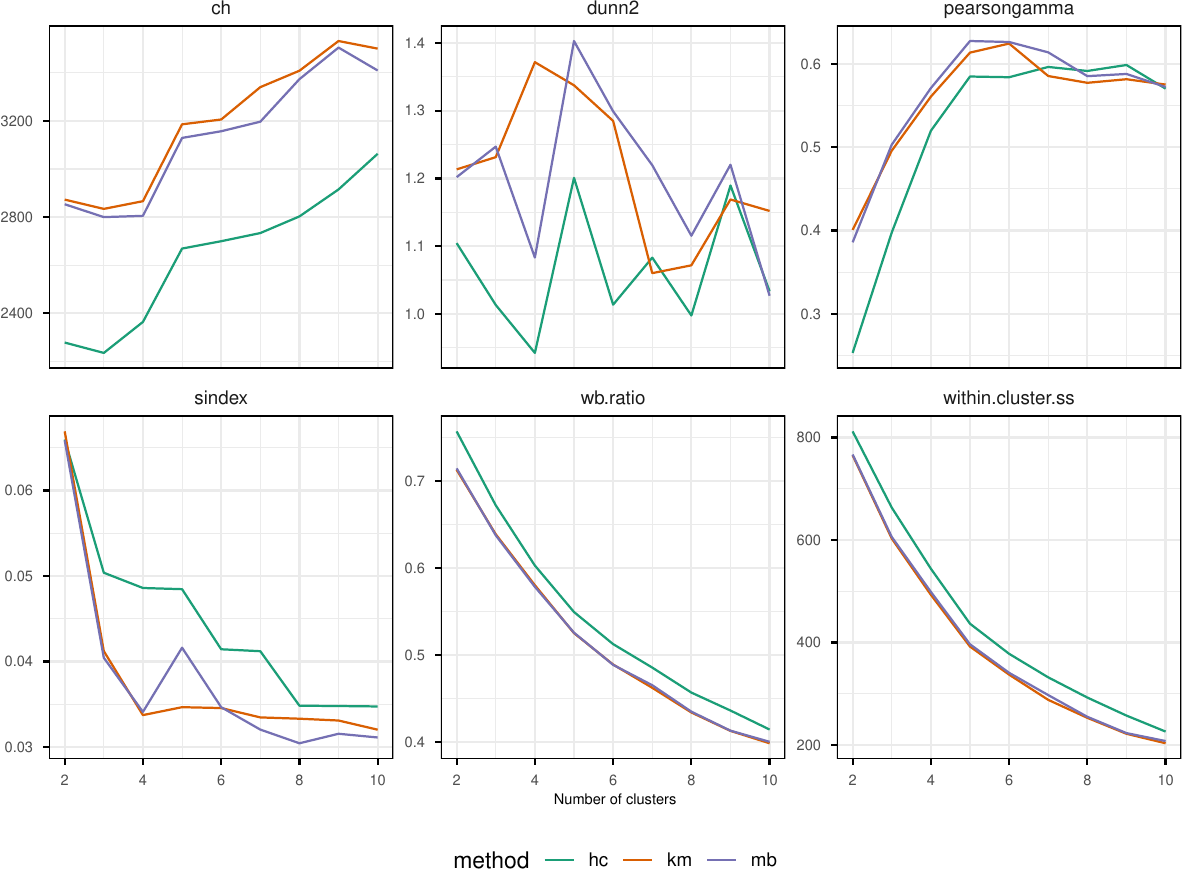} \caption{Cluster validity metrics for solutions with $2-10$ clusters obtained using $k$-means, hierarchical, and model-based clustering. Several indices consistently suggest that $4-5$ clusters provide the best balance of separation and compactness, with $k$-means performing slightly better across metrics.}\label{fig:fig-cluster-stats}
\end{figure}

Figure \ref{fig:fig-cluster-stats} shows a selection of cluster metrics for \(2-10\) clusters for each of the methods, \(k\)-means, hierarchical, and model-based. As is typical, the suggestion of the best solution varies between cluster statistics. Although the metrics differ in their preferences, several show consistent support for a \(4-5\) cluster solution. Pearson gamma (\texttt{pearsongamma}) increases sharply up to five clusters before leveling off, Calinski--Harabasz index (\texttt{ch}) increases sharply from 4 to 5 clusters and Dunn (\texttt{dunn2}) has a maximum at 5 for two methods and at 4 for \(k\)-means. All of these are interpreted as higher is better. With the other three, lower is better. WB ratio (\texttt{wb.ratio}) and within-cluster sum of squares (\texttt{within.cluster.ss}) steadily decline with number of clusters, possibly elbowing around 5 clusters. The S-index (\texttt{sindex}) is optimized at 4 clusters for \(k\)-means, 3, 6 or 8 for hierarchical clustering, and 4 or 8 for model-based. Overall, \(k\)-means performs slightly better than the hierarchical and model-based clustering across most metrics and number of clusters.

\begin{figure}[!ht]
\includegraphics[width=1\linewidth,alt={The figure shows three row-wise 2-D projection plots of synthetic 4-D data for two different k-means clustering solutions: four clusters in one panel and five clusters in the other. In each panel, the horizontal and vertical axes represent linear projections of the original high-dimensional variables, and each point corresponds to a data observation. Points are colored by assigned cluster ID (four colors in the top row and five in the bottom row). In both solutions, clusters appear as roughly compact groups, but none aligns with the underlying geometric shapes in the data: each of the five true shapes is split across multiple clusters, and no cluster cleanly isolates a single shape. The four-cluster solution merges some shapes more strongly, while the five-cluster solution yields a finer partition that still fragments each shape; additional clusters beyond five would be needed to better match the original geometric structure.}]{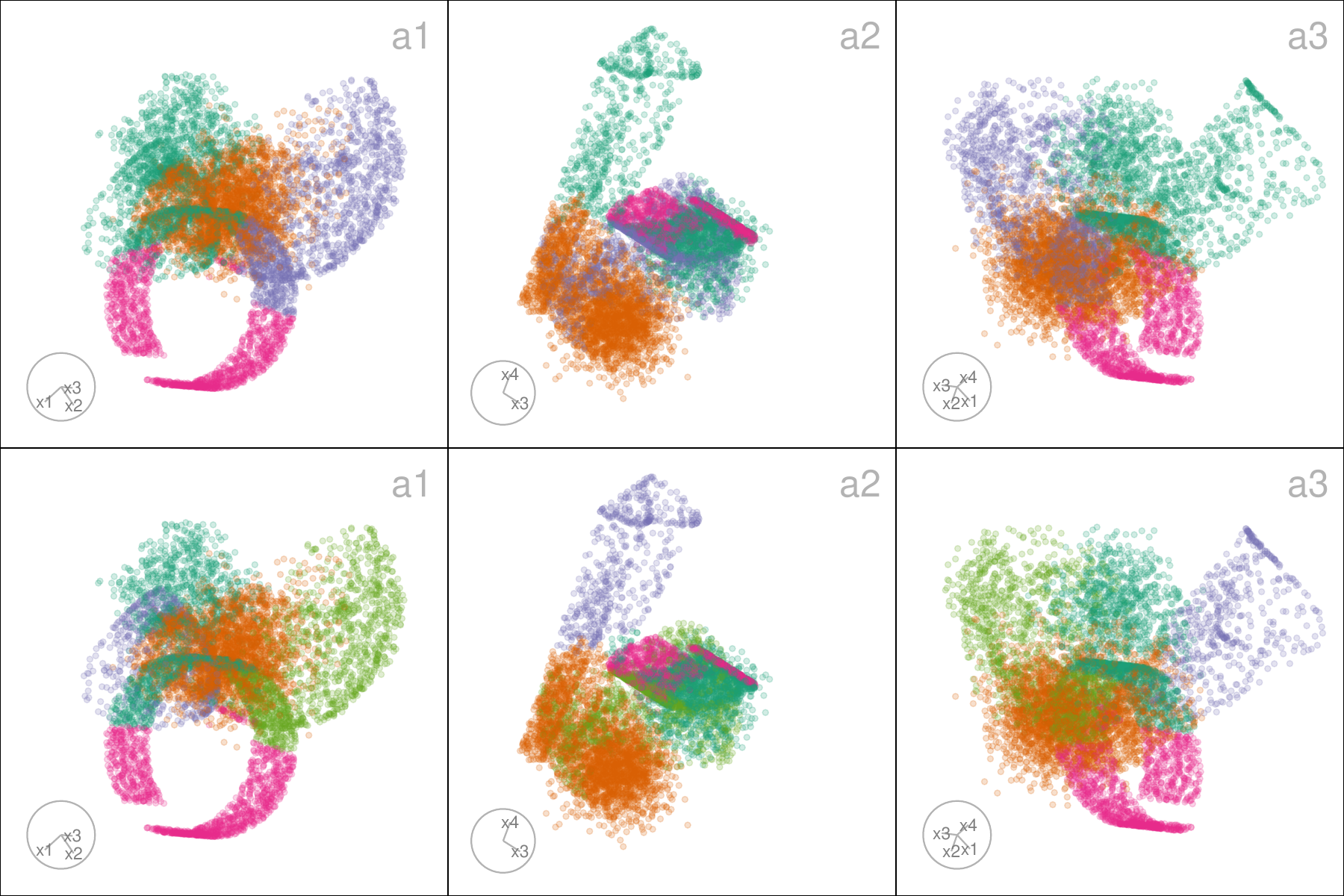} \caption{Three $2\text{-}D$ projections from $4\text{-}D$, for the five clusters data colored by the $k$-means four- (a1-a3) and five-cluster (b1-b3) solutions. The intermixing of colors within each projection reflects misclassification in both solutions, showing the difficulty of using $k$-means to capture the dataset’s nonlinear and heterogeneous shapes.}\label{fig:highd-proj-clust-algo-pdf}
\end{figure}

Figure \ref{fig:highd-proj-clust-algo-pdf} shows the four- and five-cluster \(k\)-means solutions, with cluster id used to color the points. Neither solution captures the geometric nature of the true clusters, but they are both reasonable partitions of the data. To examine either one, it is best to subset to a single cluster to view in the tour. With each solution, the five original shapes are each split by the clustering. More than 5 clusters would be needed to better capture the original shapes.

\section{Conclusion}\label{conclusion}

The \texttt{cardinalR} package introduces a flexible framework for generating high-dimensional data structures with well-defined geometric properties. It addresses an important need in the evaluation of clustering, machine learning, and DR methods by enabling the construction of customized datasets with interpretable structures, noise characteristics, and clustering arrangements. In this way, \texttt{cardinalR} complements existing packages such as \texttt{geozoo}, \texttt{snedata}, and \texttt{mlbench}, while extending the scope to higher dimensions and more complex shapes.

The included structures cover a wide range of diagnostic settings. Branching shapes facilitate the study of continuity and topological preservation, the S-curve with a hole allows investigation of incomplete manifolds, and clustered spheres assess separability on curved surfaces. The Möbius strip introduces challenges from non-orientable geometry, while gridded cubes and pyrholes test spatial regularity and clustering in sparse, non-convex regions.

These structures are designed to support not only algorithm diagnostics, but also teaching high-dimensional concepts, benchmarking reproducibility, and evaluating hyper-parameter sensitivity. By allowing users to adjust dimensionality, sample size, noise, and clustering properties, the package promotes transparent experimentation and comparative model evaluation. Together, these capabilities make \texttt{cardinalR} a versatile tool for generating interpretable, high-dimensional datasets that advance research, teaching, and evaluation of data-analytic methods.

Future extensions of \texttt{cardinalR} may include biologically inspired or application-driven data structures that would further broaden its utility in domains such as bioinformatics, forensic science, and spatial analysis.

\section{Acknowledgements}\label{acknowledgements}

The source material for this paper, including the full datasets and figures, is available at \href{https://github.com/JayaniLakshika/paper-cardinalR}{github.com/JayaniLakshika/paper-cardinalR}. This article is created using \CRANpkg{knitr} \citep{yihui2015} and \CRANpkg{rmarkdown} \citep{yihui2018} in R with the \texttt{rjtools::rjournal\_article} template. These \texttt{R} packages were used for this work: \texttt{cli} \citep{gabor2025}, \texttt{tibble} \citep{kirill2023}, \texttt{gtools} \citep{gregory2023}, \texttt{dplyr} \citep{hadley2023}, \texttt{stats} \citep{core2025}, \texttt{tidyr} \citep{hadley2024}, \texttt{purrr} \citep{hadley2025}, \texttt{mvtnorm} \citep{alan2009}, \texttt{geozoo} \citep{barret2016}, and \texttt{MASS} \citep{venables2002}.

\bibliography{paper-cardinalR.bib}

\address{%
Jayani P. Gamage\\
Monash University\\%
Department of Econometrics and Business Statistics, VIC 3800 Australia\\
\url{https://jayanilakshika.netlify.app/}\\%
\textit{ORCiD: \href{https://orcid.org/0000-0002-6265-6481}{0000-0002-6265-6481}}\\%
\href{mailto:jayani.piyadigamage@monash.edu}{\nolinkurl{jayani.piyadigamage@monash.edu}}%
}

\address{%
Dianne Cook\\
Monash University\\%
Department of Econometrics and Business Statistics, VIC 3800 Australia\\
\url{http://www.dicook.org/}\\%
\textit{ORCiD: \href{https://orcid.org/0000-0002-3813-7155}{0000-0002-3813-7155}}\\%
\href{mailto:dicook@monash.edu}{\nolinkurl{dicook@monash.edu}}%
}

\address{%
Paul Harrison\\
Monash University\\%
MGBP, BDInstitute, VIC 3800 Australia\\
\textit{ORCiD: \href{https://orcid.org/0000-0002-3980-268X}{0000-0002-3980-268X}}\\%
\href{mailto:paul.harrison@monash.edu}{\nolinkurl{paul.harrison@monash.edu}}%
}

\address{%
Michael Lydeamore\\
Monash University\\%
Department of Econometrics and Business Statistics, VIC 3800 Australia\\
\textit{ORCiD: \href{https://orcid.org/0000-0001-6515-827X}{0000-0001-6515-827X}}\\%
\href{mailto:michael.lydeamore@monash.edu}{\nolinkurl{michael.lydeamore@monash.edu}}%
}

\address{%
Thiyanga S. Talagala\\
University of Sri Jayewardenepura\\%
Department of Statistics, Gangodawila, Nugegoda 10100 Sri Lanka\\
\url{https://thiyanga.netlify.app/}\\%
\textit{ORCiD: \href{https://orcid.org/0000-0002-0656-9789}{0000-0002-0656-9789}}\\%
\href{mailto:ttalagala@sjp.ac.lk}{\nolinkurl{ttalagala@sjp.ac.lk}}%
}

\end{article}

\end{document}